\documentclass[prd,aps,nofootinbib,showkeys,preprintnumbers]{revtex4-2}
\usepackage{graphicx}
\usepackage{bbold}
\usepackage{tikz}
\usetikzlibrary{shapes.misc,calc,patterns,angles,quotes}
\usepackage{slashed}
\usepackage{amsmath}
\usepackage{amssymb}

\newcommand{\bea}{\begin{eqnarray}}
\newcommand{\eea}{\end{eqnarray}}
\newcommand{\be}{\begin{equation}}
\newcommand{\ee}{\end{equation}}

\newcommand{\cb}{\color{black}}

\newcommand{\cbl}{\color{black}}
\usepackage[colorinlistoftodos]{todonotes}
\usepackage{appendix}

\usepackage{moresize}
\newcommand{\bL}{\begin{Large}}
\newcommand{\eL}{\end{Large}}
\newcommand{\cm}{\color{black}}

\newcommand{\cPT}{\ensuremath{\mathcal{PT}}}

\def\eq{\eqref}
\newcommand{\cL}{\ensuremath{\mathcal{L}}}
\newcommand{\cZ}{\ensuremath{\mathcal{Z}}}
\newcommand{\cP}{\ensuremath{\mathcal{P}}}
\newcommand{\cT}{\ensuremath{\mathcal{T}}}
\newcommand{\cC}{\ensuremath{\mathcal{C}}}
 
\newcommand{\cCPT}{\ensuremath{\mathcal{CPT}}}

\begin{document}

\preprint{\leftline{KCL-PH-TH/2024-{\bf 06}}}

\title{Chern-Simons gravity and $\cPT$ Symmetry}
\medskip
\author{N E Mavromatos$^{a,b}$}
\author{Sarben Sarkar$^b$}
\medskip 
\affiliation{$^a$Physics Division, School of Applied Mathematical and Physical Sciences, National Technical University of Athens, Zografou Campus, Athens 157 80, Greece}
\affiliation{$^b$Theoretical Particle Physics and Cosmology Group, Department of Physics, King's College London, London, WC2R 2LS, UK}

\begin{abstract}
\color{black} This paper considers the possibility that, starting from a relativistic Hermitian quantum field theory in the ultraviolet (UV) regime, and applying a non-perturbative renormalization-group (RG) flow, we arrive at a situation where 
there are infrared (IR) singularities in the RG flow of couplings. The latter can be resolved by assuming that the theory can have a phase described by a related non-Hermitian $\cPT$-symmetric modification in the IR. \color{black} The UV-to-IR (Hermitian-to-$\cPT$-symmetric) transition can occur in a single renormalization-group flow of the pertinent couplings, as demonstrated in concrete examples. \color{black} When embedded in a gravitational setting  such a transition can lead to a repulsive gravity phase. \color{black}
If there is a  RG flow to a repulsive \cPT -symmetric gravity, then this would be an alternative to dark energy. The discussion here is presented in the context of a string-inspired Chern-Simons gravitational effective action, which involves a pseudoscalar (axion-like) field coupled to Abelian gauge fields and gravity; it may also hold more generally in gravity with torsion. 
The validity of such a scenario in realistic theories might alleviate the need for de Sitter phases in the current epoch of  cosmological evolution, thus avoiding their associated conceptual and technical complications.\color{black}

\end{abstract}


\maketitle

\section{Introduction and Motivation \label{s1}}

\cm This paper proposes a novel scenario for the observed acceleration of the universe. Rather than attributing
this acceleration to a positive cosmological constant or a dark-energy-dominated era, we interpret it instead as a
transition of the universe at large length scales to a new phase where the dynamics is described by a non-Hermitian
\cPT -symmetric gravitational theory \cite{PT2} and the gravitational interaction is repulsive. \cbl
NonHermitian \cPT-symmetric field theories~\cite{R17,qft1a,qft1b,qft1c,qft1d,qft1e,qft2,qft3,qft4,qft5,miltqed,liouv,Alexandre:2020bet,Felski:2020vrm,Chernodub,Felski:2021bdg} are 
field theory extensions of \cPT-symmetric quantum mechanics~\cite{PTqm1,PT2,PT,milt1,most,most2,most3}, which has a plethora of applications, both, experimental~\cite{Ashida:2020dkc,PTappl} and theoretical. 
 \cP ~is a linear operator (such as parity) and  \cT~is an anti-linear operator (such as time-reversal). A quantum mechanical system with \emph{unbroken} \cPT symmetry has a completely real spectrum which leads to unitary dynamics~\cite{PTqm1,PT2,most2} . 
The antilinearity, rather than Hermiticity, that characterises \cPT-symmetric theories, provides a novel guiding principle for quantum theory~\cite{maninnerPT,antilin}.  \cPT- symmetric field theories are effective field theories, which may describe aspects of Beyond-the-Standard Model physics (BSM), including massive neutrinos~\cite{R3.1, R3.2, R3.3, R3.4, R3.5, R3.6, R3.7, R3.8, R3.9, R3.10, R3.11, R3.12, R3.13, R3.14, R3.15,AB,neu1,neu2,Alexandre:2020bet}.\cm The role of \cPT-symmetric formulations changing repulsive forces to attractive forces and vice versa has a  history, which was instigated by a classic paper by Dyson on the divergence of
perturbation theory in quantum electrodynamics (QED)~\cite{R1}, He argued that the weak-coupling perturbation
expansion in QED, which is a formal series in powers of the fine-structure constant $\alpha \sim 1/137 $
, is \emph{divergent}. Dyson
argued that since the classical Coulomb force is proportional to $\alpha$, if the sign of $\alpha$ were changed by replacing the
electric charge $e$ by $ie$, the repulsive force between two electrons  would become attractive. Thus, there
is an \emph{abrupt change} in the nature of the physics at $\alpha = 0$. The possibility of classical repulsive $\cPT$-symmetric gravity was noted in~\cite{PT2} at a speculative level but with no concrete proposal.   Clearly gravity is attractive at everyday scales but at very large scales it is not. One way of understanding scales is through the renormalisation group in quantum theory.  In this article we consider a \emph{gravitating system} that exhibit singularities in their phase diagrams for large length scales. \cbl 

Our aim in this work is to pursue \cm using nonperturbative methods  a \cbl  potential gravitational aspect of \cPT-symmetric quantum physics related to the observed accelerated expansion of the universe \cite{Planck,Zhao2017-ip}. \cm We start from a Hermitian theory in the UV and see how a non-Hermitian \cPT-symmetric theory can emerge in the IR  (infrared) . Our methods are necessarily approximate, but such behaviour has been demonstrated rigorously in two dimensional minimal integrable models \cite{Lencses:2022ira,Castro-Alvaredo:2017udm}. In particular, we shall conjecture that a \cPT-symmetric phase of gravity \cite{PT2}, in which the gravitational interaction appears as repulsive,  arises within Chern-Simons (CS) gravity~\cite{jackiw,Alexander:2009tp} inspired from string theory~\cite{str1,str2,pol1,pol2,kaloper}. This phase is due to a specific singular behaviour \footnote{\cm Singular nonanalytic behaviour occurs generally due to the divergent nature of weak coupling perturbation theory.\cbl} under  nonperturbative renormalization-group (RG) flows of specific couplings in the model, related to the interactions of axionic degrees of freedom,  known~\cite{kaloper,Svrcek:2006yi} to characterise string theory spectra~\cite{str1,str2,pol1,pol2}, with CS terms; such terms arise in the theory as a result of the Green-Schwarz mechanism~\cite{GS} for the cancellation of gauge and gravitational anomalies in the extra-dimensional spaces of compactified strings. Such systems involve axion (pseudoscalar) degrees of freedom coupled to topological Pontryagin anomalous densities of both gravitational and gauge nature. For simplicity we consider the \cm restricted  case of Abelian gauge sectors and the absence of gravitational anomalous terms, i.e.  axion electrodynamics,. For this case,  we show that the singularities in the phase diagram of the quantum  theory lead to inconsistencies that can be remedied (by an appropriate phase transition) on replacing the Hermitian theory by a nonHermitian but \cPT-symmetric  version. \cbl 

We shall consider string effective theories in which gravitational and gauge CS anomalous terms survive in the (3+1)-dimensional spacetime after string compactification. 
The flat spacetime  limit of such constructions leads to an axion electrodynamics model, once the gauge sector CS terms are restricted to Abelian U(1) (electromagnetic) gauge groups. 
On coupling the \cb axion electrodynamical \cbl system to dynamical gravity, we  conjecture that the \cPT-symmetric non-Hermitian IR phases may correspond to repulsive gravitational interactions at large (cosmological) scales. We reinterpret existing works in the literature on non-perturbative renormalisation group (RG) studies of such axion-electrodynamics models~\cite{Eichhorn:2012uv}, in terms of potential phase transitions of the Hermitian system to a non-Hermitian but \cPT-symmetric one. The transition occurs at the singularities of the RG beta functions in the infrared (IR) region, which such systems exhibit. Should such a phase be valid in realistic cosmologies, it might describe the observed accelerated expansion of the universe at late eras, thus avoiding the technical and conceptual issues associated with de Sitter vacua \cite{Agmon:2022thq}, especially from the point of view of string theory.  Positive cosmological-constant 
(de Sitter) gravitational backgrounds appear in conflict with both perturbative~\cite{scat1,scat2} and non-perturbative 
string theory frameworks~\cite{swamp1,swamp2,swamp3}. The presence of (observer-dependent) de Sitter horizons, prevents the existence of a well-defined scattering S-matrix, due to the lack of asymptotic ``in'' and ``out'' states. Also the  failure of the decoupling of infinite towers of massive string states at specific points of the moduli spaces of string theory in de Sitter backgrounds, leads to the incompatibility of such backgrounds with a consistent embedding in a  quantum gravity setting. Thus the latter models appear to belong to  the so-called ``swampland regions'' \cite{swamp3} of quantum gravity theories. Therefore such frameworks allow  only quintessesnce models for dark energy as the cause of the currently observed accelerating expansion of the universe.

 \cm $\cPT$ symmetric   quantum mechanics poses some technical issues because it is non-Hermitian\cbl. It is usually treated using  canonical quantisation and Hilbert space methods \cite{PT2}. \cm For unitarity the \cbl Hilbert space requires an inner product which is different from the Dirac inner product used for Hermitian theories. \cm Even in quantum mechanics (with a finite number of degrees of freedom) it is not usually possible to calculate this inner product exactly. \cbl For an infinite number of degrees of freedom the construction of the inner product can only be done perturbatively for ``realistic'' (rather than toy) field theories. \cm The resultant perturbative quantum theory may be  highly singular \cite{Bender_2006} and so not useful.\cbl \footnote{\cm When $\cPT$ symmetry is broken, such a construction is not  possible.\cbl} Quantum theory, however, has an alternative formulation in terms of path integrals and associated  functional methods in general spacetime dimensions $D$, \cm  which can alleviate the above difficulties \cbl . 
 The partition function (or generating functional) is expressed as a path (functional) integral \cite{Rivers:1987hi} and serves as a \emph{starting point} for both perturbative and nonperturbative calculations of \cm Schwinger functions  and beta functions \cbl.  The path integral (in analytic approaches) can be defined semiclassically for \emph{weak coupling}. For a theory to be defined this path integral needs to exist \cite{R2}, whether in a Hermitian  or $\cPT$ symmetric context. In the semiclassical framework paths are chosen to be the functional \emph{analogue} of steepest-descent paths \cite{Witten:2010cx,R32w}
 (terminology from $D=0$) determined by Picard-Lefschetz theory . In the language of fixed points  suited for  $D=0$,  the steepest descent paths connect the trivial fixed point to non-trivial saddle points. The imaginary contributions, which arise from the non-trivial saddle points, are cancelled by the imaginary part of the Borel sum of the perturbative fixed point. This is the crucial nonperturbative physics of $\cPT$ symmetry \cite{R2bb,ABS} present at weak coupling, which is harder to unravel using canonical methods. In this approach the Feynman diagrams follow from gaussian evaluations around the trivial fixed point. \cm  Although weak coupling analysis in simple models does suggest the possibility of  flows between Hermitian and \cPT~-symmetric fixed  points \cite{Croney:2023gwy}, in our gravitational model we require a strong coupling nonperturbative analysis, which is possible using functional methods. \cbl

 The functional approach  has been studied  at length recently \cite{Saueressig:2023irs,Reichert:2020mja,R3a,Croney:2023gwy,ABS,R3a}.  It leads to Schwinger-Dyson equations \cite{qft1a,R3c}  and the functional renormalisation group (FRG) equations \cite{Ai:2022olh,Wipf:2013vp}, which  are compatible with $\cPT$ symmetry. \cm The role of the inner product in this approach  can be seen clearly in a \cPT~symmetric  theory for a pseudoscalar field $\phi$ with a potential $U(\phi)$ \cbl. The partition function $Z[j]$ (in the presence of a source\footnote{To formally maintain $\mathcal{PT}$ symmetry in the presence of a nonzero source, the source can be coupled to $i \phi $~\cite{Ai:2022olh}.} $j(x)$) can be expressed in two different ways~\cite{Jones:2009br}, $Z_{1}[j]$ and $Z_{2}[j]$, a time ($T$) ordered vacuum expectation value~\cite{Rivers:1987hi}; so
\be
\label{E39}
Z\left[j\right]=Z_{1}[j]=\int D\phi\exp\left(iS\left[\phi\right]-\int  j\left(x\right)\phi\left(x\right)\right)
\ee
and
\be
\label{E40}
Z\left[j\right]=Z_{2}[j]= \left\langle 0 \left| \, \eta T\left(\exp\left[-\int dx \, j(x) \, \phi(x)\right]\right) \right| 0 \right\rangle.
\ee
with $\eta$ as the time-independent metric of canonical $\cPT$ theory. The representation of $Z[j]$ by $Z_{2}[j]$ is known as the Symanzik construction~\cite{Rivers:1987hi}. The Schwinger-Dyson equations (SDE) obtained from $Z_{1}[j]$ and $Z_{2}[j]$ are known to coincide \cite{R3a}. Solutions of the SDE give the content of the quantum theory. On taking the Legendre transformation of the generating functional  $Z_{1}[j]$, we can derive  the Wetterich  equation used in the functional renormalisation group flow~\cite{wetterich1993exact}. \cbl. The fact that nonperturbative renormalisation can lead from a Hermitian theory to $\cPT$ symmetric theory is the source of understanding the Lee model \cite{R6.1,R5}, which before the \cPT -symmetric interpretation was regarded (for decades) as full of ghosts. \cbl

\subsection{Non-analyticity and the rise of \cPT-symmetric theories} \label{NA}

Motivated by  these considerations, \cm we give arguments in favour of \cPT-symmetric gravitational theories which may offer alternative explanations of the current-era cosmological data. 
Our investigation is prompted by intriguing developments in a field theoretic study of  gravitational axion phenomenology and dynamical mass generation~\cite{R3.13,Alexandre:2020tba,R3.14,R3a,R3c}. We noticed a RG flow~\cite{R3a} from Hermitian values of the coupling to those of a non-Hermitian but $\cPT$-symmetric version of the field theory in a one-loop analysis. This might be interpreted as a dynamical appearance of \cPT ~symmetry, as  in the nonHermitian Nambu-Jona-Lasinio model, for instance, where the nonHermiticity is claimed to appear spontaneously~\cite{Chernodub}.\cbl
We further examine the robustness of these findings by working with beta functions with non-zero $\epsilon$ (where spacetime dimension $D=4-\epsilon$) and by working to three loops in the Yukawa coupling and two loops in the quartic coupling~\cite{R10, Croney:2023gwy}. The idea that the effects of renormalisation can lead from a Hermitian theory to a $\cPT$ symmetric theory is first demonstrated in a different context in the Lee model \cite{R6.1, R5}, a solvable toy model. The concepts of $\cPT$ symmetry are most  developed in the context of quantum mechanics however \cite{PTqm1}. $\cPT$ symmetry is realised not just in the form of the Hamiltonian, but implies also a different\emph{phase} of the system \cite{R17,PhysRevLett.130.250404,ABS} with an energy spectrum which cannot be obtained continuously from a Hermitian theory by analytic continuation in the  couplings of the theory \cite{R6,Simon:1970mc,Simon1991FiftyYO}. \cm Although the $\cPT$ phase is  understood for the quantum mechanical anharmonic oscillator (with negative coupling), it can only be seen approximately in higher dimensional quantum field theoretic systems through, for example,  singularites in renormalisation group flows. \cbl

First let us consider RG flows in the simpler context of chiral Yukawa interactions of axion fields $\phi(x)$ coupled to fermion fields $\psi(x)$ with coupling constant $g$ \cite{R3.13,R3c,R3a}, 
\begin{align}\label{e1}
\cL_{\rm Yukawa}=ig\,\phi\,\overline\psi\,\gamma^5\,\psi.
\end{align}
The one-loop RG $\beta$ function of $g$ in $D$ spacetime dimensions as $\epsilon\to 0^+$
leads to\footnote{An analysis with respect to the coupling $g$ rather than $g^2$ was made in Refs.~\cite{R3a,R3c} but in this paper we restrict our attention to RG flows of $g^2$. In ~\cite{Croney:2023gwy} the analysis is extended to higher loops where the renormalization group flows show the same qualitiative features.}
\begin{align} 
\label{e2}
\frac{d}{dt}(g^2)=\frac{5}{8\pi^2}(g^2)^2-\epsilon g^2,
\end{align}
where $d/dt\equiv\mu\,d/d\mu$, and $\mu$ is a transmutation mass scale.
The solution to \eqref{e2} with $\epsilon t \ll 0$ is 
\be\label{e3}
g^2\approx -\frac{1}{C+\frac{5}{8\pi^2}t}\,,
\ee
where $C$ is an integration constant. \cb This calculation is only heuristic but persists at two loops. It is admittedly perturbative so we cannot  conclude that it will definitely persist as a  nonpertubative effect \cite{R3a}. \cbl For Hermitian couplings [where $g$ in \eqref{e1} is real] at $t=0$ and $C<0$, we see from \eqref{e3} that $g^2$ increases with increasing $\mu$ until $g^2$ hits a pole at finite $t=t_p= 8\pi^2|c|/5>0$ \cite{Chanowitz:2000ka}. This is similar to the Landau pole in Abelian U(1) gauge theories and may be a sign of new physics. In terms of an infrared (IR) cutoff $m$, $t=\log (\mu/m)$. Here, $t=0$ corresponds to an IR (large-distance) regime and $t\to\infty$ ($\mu\gg m$) corresponds to an ultraviolet (UV) (short-distance) fixed point.

Perturbation theory (\cm at this order \cbl) breaks down at the pole, and one must decide how to go around the pole. Formally, for values of $t>t_p$ the theory becomes non-Hermitian because $g^2<0$. However, this theory is $\cPT$-{\it symmetric} and {\it asymptotically free} \cite{PhysRevD.82.085013} because $g^2\to 0^-$ as $t\to\infty$. The $\beta$ function of the Hermitian theory ends up at a pole (starting from some IR regime), while the $\beta$ function of the $\cPT$-symmetric non-Hermitian theory starts at that pole, and then exhibits asymptotic freedom in the UV. \cm Given the approximate nature of this calculation the precise nature of this singularity cannot be deduced and is unlikely to be a pole, and could even be an essential singularity, in a nonperturbative  treatment .This indicates that one could indeed go past this singulalrity to the asymptotically free non-Hermitian \cPT-symmetric version, which has the same observables \cbl.\footnote{Since the \cPT ~theory has in general a different dynamical behavior, and a different phase space than the initial theory, the word ``analytic continuation" here should be interpreted as meaning the substitution of the pertinent coupling $g$ by its purely imaginary image, $ig$, $g\in \mathbb R$. This is to be understood throughout the current article.}

A conjectural prescription based on analysis of scalar field theories in $D=1$ is given in Ref.~\cite{ABS}
that allows the analytic continuation of a Hermitian (but ill-defined) self-interacting scalar theory in $D$ spacetime dimensions to a well-defined non-Hermitian
$\cPT$-symmetric theory. The conjecture in higher $D$ (which is not proven) states that if $\cZ(\lambda)$ is the partition function of the path integral of the Hermitian self-interacting scalar theory with coupling $\lambda$, then the corresponding partition function $\cZ_\cPT(g)$ of the $\cPT$-symmetric theory with self-interaction coupling $g$, is given by
\begin{align}\label{e4}
{\rm ln}\cZ_\cPT(g)={\rm Re} \big[{\rm ln}\cZ(\lambda=-g+i\, 0^+)\big].
\end{align}

\noindent This conjecture suggests a mechanism for avoiding the aforementioned poles in the RG $\beta$ functions by analytically continuing the theory to a $\cPT$-symmetric theory. \cm However a word of caution: from the work of Bender and Wu \cite{Bender:1969si,Simon1991FiftyYO} on the anharmonic oscillator the singularity structure in the complex $g$-plane is very complicated and is full of nonanalytic singularities. The path of any analytic continuation may be tortuous.\cbl

Recently, the conjecture was applied in \cite{Romatschke} to treat the UV pole that appears in the RG $\beta$-function of the $O(N)$ scalar models at large $N$ with Hermitian self-interactions of the form (in Euclidean formalism)
\begin{align}\label{oN}
\mathcal L_{\rm int} = \frac{\lambda}{N} \, (\vec \phi^{\,T} \cdot\vec \phi)^2 \,, \quad  N \gg 1, \quad \lambda \in \mathbb R\,,
\end{align}
where $\vec\phi$ denotes the $N$-component scalar field and $T$ indicates matrix transposition \cite{Chanowitz:2000ka}. The avoidance of the pole in the conjecture of \cite{ABS} is accomplished under the condition that the physical observables of the conjectured $\cPT$-symmetric theory
in the IR and the original Hermitian theory in the UV remain the same and the poles in the $\beta$ function do not affect the physical observables.\footnote{There are unresolved subtleties in defining a $\cPT$-symmetric path integral for $N$-component fields. Moreover, the conjecture of \cite{ABS} has been scrutinised in \cite{kamata}, where, within a WKB approximation it was argued that the conjecture is valid under certain conditions.}

The Hermitian large-$N$ theory is known to be characterized by a positive RG $\beta$-function and a Landau pole in the UV~\cite{RomONLP},
as found for the renormalised coupling \eqref{e3} of the Yukawa interaction \eqref{e1} in (3+1)-dimensional models \cite{R3a,R3c,Croney:2023gwy}. 
The $\cPT$-symmetric theory is obtained by replacing $\lambda\to-g$ in the Euclidean formalism (taking care of the Riemann sheet structure in complex $g$ shown in detail for the quantum anharmonic oscillator~\cite{Bender:1969si}), where $g$ denotes the $\cPT$-theory coupling; following the conjecture \eq{e4} of Ref.~\cite{ABS}, one analytically continues $\lambda\to-g+i\,0^+$ in the corresponding expressions of the initially Hermitian theory, such as thermodynamic quantities at finite temperature~\cite{Romatschke}.
The nonperturbative (large $N$) renormalized coupling $g_R(\mu)$ of  $\cPT$ symmetry (in dimensional regularization in $D=4-\epsilon$ dimensions, $\epsilon\to0^+$), 
exhibits a {\it negative} $\beta$ function~\cite{Romatschke}, 
\begin{align}\label{bON}
\beta=\frac{d}{d\,\ln\mu} g_R(\mu)=-\frac{8\pi^2 }{\ln^2(\frac{\mu^2}{\Lambda_c^2})}=-\frac{g_R^2(\mu)}{2\pi^2}<0, 
\end{align}
where $\mu$ is a transmutation mass scale. The scale $\Lambda_c$ denotes the scale at which the running coupling $g_R(\mu)$ develops a pole as $\mu$ increases from $\mu=\Lambda_c$ towards the UV:
\begin{align}\label{polegR}
g_R(\mu)=\frac{4\pi^2}{\ln(\frac{\mu^2}{\Lambda_c^2})}\,.
\end{align}
Notably, the expressions \eqref{bON} and \eqref{polegR} are exact in the large-$N$ limit. As in the Yukawa case \eq{e3}, the renormalised coupling of $\cPT$ symmetry becomes small in the UV; that is, it exhibits {\it asymptotic freedom}. 
 
In Ref.~\cite{Grable} the authors apply the conjecture of \cite{ABS} to a $(4-\epsilon)$-dimensional theory, $\epsilon\to0^+$, with fermionic self-interactions (the large-$N$ Gross-Neveu model of a four-fermion interaction among $N$-component fermions):
\begin{align}\label{GN}
\cL_{\rm int}=\Big(\frac{\alpha}{N}\Big)^{\frac{1}{D-1}}\, 
\Big(\overline \psi_f\,\psi_f \Big)^{\frac{D}{D-1}}\,,
\end{align} 
where $\psi_s^T=(\psi_2,\psi_2,\dots,\psi_N)$, represents the $N$-component fermion fields and $\alpha$ is the coupling. The Hermitian versions of such theories are asymptotically free in the UV and exhibit IR poles in their RG $\beta$ functions (a situation opposite to the Yukawa coupling of \cite{R3a,R3c,Croney:2023gwy} examined above. The avoidance of the $\beta$-function pole is similar). 
 
The \cm switch \cbl in Ref.~\cite{ABS} to the counterpart $\cPT$-symmetric non-Hermitian theories in the IR leads to interesting phase transitions in a finite-temperature analysis. Indeed, the renormalized 
coupling of the Hermitian theory $\alpha(\mu)$ has a negative $\beta$-function 
\begin{align}\label{GNb}
\beta_{\alpha}=\frac{d\,\alpha_R (\mu)}{d\,\ln\mu}=-\frac{\alpha_R^{2}(\mu)}{2\pi^2}\,,
\end{align}
which is opposite in sign from the corresponding $\beta$-function of the large-$N$ scalar theory ~\cite{Romatschke}. Its solution leads to the existence of a pole at some cutoff scale $\Lambda_c$ in the IR, and asymptotic freedom in the UV ($\mu/\Lambda_c \to 0^+$):
\begin{align}\label{GNpolegR}
\alpha_R(\mu)=4\pi^2/\ln(\mu^2/\Lambda_c^2)\,.
\end{align}

On the other hand, for large $N$ the nonperturbatively  renormalized coupling of the $\cPT$ theory, 
$\alpha_R^\cPT(\mu)$, obtained by the analytic continuation procedure of Ref.~\cite{ABS}, has a positive $\beta$ function, which decays into the IR regime,
\begin{align}\label{GNbPT}
\beta_\alpha^\cPT=\frac{d}{d\ln\,\mu}\,\alpha_R^\cPT(\mu)=\frac{\alpha_R^\cPT(\mu)^2}{2\pi^2}\,\quad\Rightarrow\,\quad\alpha_R^\cPT(\mu)=-
\frac{4\pi^2}{\ln(\frac{\mu^2}{\Lambda_c^2})}\,,\quad\mu<\Lambda_c\,.
\end{align}
 It is hoped that such simple models may contribute to a nonperturbative understanding of the rich phase structure such systems may posses.\footnote{The thermodynamical analysis of Ref.~\cite{Grable} reveals also an interesting first-order phase transition at the pole, separating stable, meta-stable, and unstable phases.}
\vskip .3cm
\subsection{Main Theme and Structure of the article}\label{sec:Theme}

In this article we consider \emph{gravitating systems} that exhibit singularities in their phase diagrams for large length scales. These systems involve axion degrees of freedom coupled to topological Pontryagin anomalous densities of both gravitational and gauge nature. These couplings are related. Such self-gravitating axion systems have a natural interpretation as low-energy systems of \color{black} (3+1)-dimensional \emph{string-inspired} gravity (after compactification)\color{black}, where the axion plays the r\^ole of the dual of the field strength of the Kalb-Ramond (KR) spin-one field of the massless gravitational multiplet of the string. \color{black} This KR axion is the so-called string-model independent axion, which seems to be present in all string models. In addition, in string theory there are additional axions, coming from compactification, which depend on the underlying microscopic string theory model, and they are not part of the string massless gravitational multiplet~\cite{Svrcek:2006yi}. \color{black}
We can simplify to the case of Abelian gauge sectors with no gravitational anomalous terms. This simplification allows us to make use of results from axion electrodynamics \cite{Eichhorn:2012uv}. We demonstrate that there are singularities in the phase diagram of the theory, which lead to inconsistencies that can be remedied (by appropriate phase transitions) involving a replacement of the Hermitian theory by a non-Hermitian  \cPT-symmetric  version. 
\color{black} As we shall discuss in this article, when embedded in curved space time, such models lead to gravitational models with negative values of the Newton constant. This feature should be contrasted with the usual 
(in the context of (weak) quantum Einstein Gravity) running of the Newton constant (asymptotic safety approach), where, as we shall see in the next section, the flow never leads to repulsive gravity phase~\cite{reuter,Hamber,Hamber92,Hamber99}\color{black}.

The KR (or gravitational) axions may also be viewed as \color{black} (3+1)-dimensionl duals to \color{black} totally antisymmetric torsion in the geometry of the string-inspired effective theory \cite{Gross:1986mw,kaloper}, at least up to quartic order of target-spacetimes derivatives in a Regge slope $\alpha^\prime$ expansion~\cite{str1,str2,pol1,pol2}.
Such a link could also lead to the conjecture that the aforementioned behavior may characterize more general gravitational theories with torsion \cite{hehl,shapiro,Hammond:2002rm,PhysRevD.104.084067,iorio}, such as Einstein-Cartan theory \cite{Cartan:1938ph}\cm 
\, and Chern-Simons gravity \cite{jackiw} \cbl and not just standard Einstein gravity.

\color{black} It is important to contrast our approach with a quantum field theoretic approach with mixed repulsive and attractive interactions. It is possible that a renormalization group flow of the respective couplings might allow a trajectory between basins of attraction of fixed points \cite{Hollowood:2009eh} which
does not require any singular behaviour. 
For instance, in cases  where the relevant potential contains the difference between two Yukawa potentials, one could in principle define (typically violating spectral positivity) an appropriate combined running coupling such that the total force switches
from attractive to repulsive. 
 This case is not in general associated with the existence of a singular point of the renormalization-group flow, unlike our case discussed above, 
because the (renormalized) couplings of the two Yukawa interactions typically have independent fixed points. Our case is not simply 
a non-Hermitian quantum field theory with attractive and repulsive
interactions in the same channel. 
As we shall argue in section \ref{sec:repgrav}, the change of the sign of the gravitational interaction is \emph{induced} by the existence of singular points of (the non-perturbative) renormalization-group flows of the axion-electrodynamics part of the curved-space string-inspired CS effective action in \eqref{sea3}, which necessitates a non-trivial jump to the \cPT ~symmetric framework, in order to bypass such a singularity ({\it cf.} section \ref{s3}). \color{black}

This paper is organized as follows: 
In section \ref{sec:asymptsafety}, we briefly review some results from the asymptotic safety approach to quantum gravity, as regards  UV versus IR properties of the running of the gravitational constant with the renormalization group (RG) scale; this running will be correlated with the running of the coupling in our axion electrodynamics.  
In Sec.~\ref{s2} we give details of the superstring-inspired gravitational model and explain the origin of the axion and its anomalous gauge and gravitational couplings in this context. 
In Sec.~\ref{s3}, 
based on a reduction of the gauge structure of the string-inspired model to Abelian electromagnetic interactions only, we discuss the 
nonperturbative renormalization of the coupling in (a flat-spacetime) axion electrodynamics, the prototypical Hermitian model. The RG flow shows a singular  infrared behaviour which lies near the nontrivial infrared fixed point of the model. Although this singularity is much more complicated than, for example, a pole (noted in an earlier example),
we propose to bypass such a singular flow behavior by replacing the theory by its non-Hermitian \cPT-symmetric counterpart, which has a trivial IR fixed point but a nontrivial UV fixed point. 
In Sec.~\ref{sec:repgrav}, we connect this flat spacetime model to our (string-inspired) gravitational theory (in a dynamically curved spacetime) and provide arguments on how the analytically extended \cPT ~version of the model in the infrared results in a phase transition to a repulsive-
gravity phase at large scales. We also discuss potential subtleties concerning the validity of this conjecture in the quantum gravity sector, which itself has  a RG running of the gravitational coupling \cite{Reichert:2020mja} (Newton's ``constant''). Our conjecture rests on the presence of the (singularly) running axion coupling; the gravitational coupling changes sign at the infrared and leads to a repulsive-gravity phase at large scales. Our conclusions and outlook are given in Sec.~\ref{sec:concl}.

\section{Asymptotic Safety and Quantization of Gravity}\label{sec:asymptsafety}

The quantization of Einstein gravity as an ordinary field theory leads to the well-known asymptotic safety of Einstein (3+1)-dimensional gravity \cite{reuter,Reuter_Saueressig_2019}, and the existence of a UV fixed point for the running Newton constant \cite{reuterweyer, Hamber}. For our purposes we follow the treatment in \cite{Hamber}, and references therein (in particular \cite{Hamber92, Hamber99}), which is closer in spirit to the work here. The approach is based on a quantization of gravity 
on a Euclidean lattice, using a discrete higher-derivative action whose naive classical continuum limit corresponds to the action
\begin{align}\label{continact}
\mathcal S=\int d^4x \, \sqrt{-g}\,\Big[-\frac{1}{2\kappa^2} R+b \, R_{\mu\nu\rho\sigma}\,R^{\mu\nu\rho\sigma}+\frac{1}{2}(a-4b)\, C_{\mu\nu\rho\sigma}\, C^{\mu\nu\rho\sigma}+\Lambda \Big]\,,
\end{align}
where $a$ and $b$ are dimensionless constants, $C_{\mu\nu\rho\sigma}$ is the Weyl tensor, and $\Lambda$ is a cosmological constant term. The higher-derivative terms in the action help to avoid unboundedness (from below) of the Einstein-Hilbert action on the lattice \cite{Kellett2021-ua,Gibbons1978-tz}. For string-inspired quantum theories of gravity, such higher-derivative additions are natural. 

According to that approach to gravity quantization, the RG running of Netwon's constant $\rm G=8\,\pi\,\kappa^2$ in the vicinity of the UV fixed point is described by
\begin{align}\label{rgG}
{\rm G}(k^2)=8\,\pi\, \kappa^2={\rm G_c}\,\Big[1+a_0 \Big(\frac{m_{\rm IR}^2}{k^2}\Big)^{1/2\nu} + \mathcal O \Big( \Big( \frac{m_{\rm IR}^2}{k^2} \Big)^{1/\nu}\Big)\Big]\,,
\end{align}
where $k^2$ is a running mass scale, and $m_{\rm IR}^2$ is an IR mass scale, which can be taken to be of the \emph{order of today's Hubble parameter $H_0$}, if we are to explain the current-epoch acceleration of the expansion of the universe. The detailed computations of \cite{Hamber92,Hamber99} have shown that $a_0>0$ and the critical exponent $\nu=1/3$. The running Newton's coupling flows to $\rm G_c$, asymptotically for large $k \gg m_{\rm IR}$ (short distances); $\rm G_c$ can be identified with the physical value of Newton's ``constant'', measured at laboratory scales, {\it i.e.} (in units $(\hbar\, c=1$): $\sqrt{8\pi \, \rm G_c} = \sqrt{8 \pi\, \rm G_{\rm phys}} \sim M_{\rm Pl}^{-1}$, with the reduced Planck mass scale $M_{\rm Pl}= 2.4 \times 10^{18}$~GeV.

The UV behavior \eqref{rgG} is model dependent, but the approach of $\rm G$ to an UV fixed-point value $\rm G_c$ as $k^2/m^2_{\rm IR} \to \infty$ seems to be model independent~\cite{reuter}. On the other hand, the expression \eqref{rgG} is ill-behaved in the IR, $k^2/m^2_{\rm IR} \to 0^+$, where it develops a {\it pole} that needs to be {\it regulated}. Such a regularisation is highly model dependent. In 
\cite{Hamber}, the following regularization has been suggested, by means of which the pole $1/k^2$ is replaced simply by $1/(k^2 + m^2_{\rm IR})$. Under such a scheme one may write for the IR behavior of the running Netwon's coupling in the models of \cite{Hamber, Hamber92,Hamber99}
\begin{align}\label{irG}
{\rm G}(k^2) \simeq {\rm G_c} \, \Big[ 1 + a_0 \Big(\frac{m_{\rm IR}^2}{k^2 + m_{\rm IR}^2}\Big)^{\frac{1}{2\nu}} + \dots \Big]\,.
\end{align}
If this assumption is correct, this would imply the following IR (large (cosmological) distance) behavior of the running Netwon's coupling in (3+1)-dimensions:
\begin{align}\label{IRgr}
\lim_{k^2/m^2_{\rm IR}\to 0^+} {\rm G}(k^2)={\rm G}_\infty \, \Big[1 - \Big(\frac{a_0}{2\, \nu \, (1 + a_0)} + \dots \Big)\, \frac{k^2}{m^2_{\rm IR}} + \mathcal O(\frac{k^4}{m^4_{\rm IR}})\Big]\,,
\end{align}
where ${\rm G}_\infty={\rm G_c} (1 + a_0 + \dots)$, independent of the IR cutoff $m_{\rm IR}$. 

In the above formulas the UV cutoff $\Lambda_{\rm UV}$, which can be identified with the (reduced) Planck scale $M_{\rm Pl}$, appears inside the expression for $\rm G_c = \Lambda_{\rm UV}^{-2} \, \widetilde{\rm G}_c$, with $\widetilde{\rm G}_c$ a dimensionless coupling.
In this scheme for regulating the IR pole in the large-distance behavior of the renormalized Newton's coupling G, the latter is {\it always positive}, thus gravity appears as {\it attractive} during the entire RG flow from UV to IR.
In such a formalism, the addition of a positive cosmological constant $\Lambda>0$ can be interpreted as leading to an effectively repulsive gravity that leads to the current-era acceleration of the universe ~\cite{Planck}. However, in such a de Sitter Universe, the asymptotic Newtonian limit cannot be rigorously defined, and hence such an interpretation is not strictly valid ~\cite{bernabeu} unless the de Sitter phase is metastable. 
 
Bypassing this IR pole will be of concern to us below, where we argue that such a task can also be achieved by passing to a non-Hermitian but $\cPT$-symmetric version of gravity that we shall define properly within the context of CS gravity~\cite{jackiw,Alexander:2009tp} inspired by string theory \cite{str1,str2,pol1,pol2}. The nonHermitian theory will be argued to characterise a repulsive phase of 
the anomalous CS gravity.
This model requires a detailed discussion not commonly found in treatments of ordinary Einstein gravity theory, 
as we see in the next section.

\section{String-inspired Models of Chern-Simons Gravity \label{s2}}

In superstring theory ~\cite{str1,str2,pol1,pol2}, after compactification to four space-time dimensions, the bosonic ground state of the closed-string sector consists of {\it massless} fields in the so-called {\it gravitational multiplet}, which contains a spin-$0$ (scalar) dilaton $\Phi(x)$, 
a spin-2 traceless symmetric tensor field, $g_{\mu\nu}(x)$, which is uniquely identified as the  (3+1)-dimensional graviton, and a spin-1 antisymmetric tensor gauge field $B_{\mu\nu}(x)=-B_{\nu\mu}(x)$, known as the Kalb-Ramond (KR) field. In what follows, for brevity and concreteness, we set the four-dimensional dilaton field to a constant, $\Phi(x)=\Phi_0$. This fixes the string coupling 
\be\label{stringcoupl}
g_s=\exp(\Phi)= \exp(\Phi_0)~. 
\ee
There are always consistent solutions of the four-dimensional string theory with such a configuration, and this suffices for our purposes.

There is a $U(1)$ gauge symmetry of the closed-string $(3+1)$-dimensional target-space-time effective-field-theory action, associated with the KR $B$-field transformations 
\be\label{Bgauge}
B_{\mu\nu}(x) \, \rightarrow \, B_{\mu\nu}(x) + \partial_{[\mu}\theta_{\nu]}(x), \quad \mu,\nu =0,\dots 3, 
\quad \theta_\mu (x) \in \mathbb R, 
\ee
where Greek indices from now on denote space-time indices, taking on the values $0, \dots 3$, and 
the symbol $[\dots]$ denotes antisymmetrization of the respective indices.
The $U(1)$ gauge symmetry of the closed-string sector implies that the corresponding effective action will be expressed only in terms of the field strength of the $B$-field:
\be\label{KRfs}
\mathcal H_{\mu\nu\rho}(x) = \partial_{[\mu} B_{\nu\rho]}(x).
\ee
This is subject to the following Bianchi identity 
\be\label{bianchi}
 \mathcal H_{[\nu\rho\sigma\, ;\, \mu]}=\partial_{[\mu}\mathcal H_{\nu\rho\sigma]} = 0~.
\ee
From now on, we omit the space-time-coordinate arguments of the fields for brevity. The semicolon denotes covariant derivative with respect to the standard
Christoffel connection $\Gamma_{\,\,\mu\nu}^\alpha= \Gamma_{\,\,\nu\mu}^\alpha$ of the metric $g_{\mu\nu}$.   

We next make the important remark that in superstring theory anomaly cancellation requirements imply a modification of the KR field strength \eq{KRfs} by 
appropriate gauge (Yang-Mills (Y)) and Lorentz (L) (gravitational) CS terms (Green-Schwarz mechanism)~\cite{str2}
\begin{align}\label{csterms}
\mathbf{{\mathcal H}} &= \mathbf{d} \mathbf{B} + \frac{\alpha^\prime}{8\, \kappa} \, \Big(\Omega_{\rm 3L} - \Omega_{\rm 3Y}\Big),  \nonumber \\
\Omega_{\rm 3L} &= \omega^a_{\,\,c} \wedge \mathbf{d} \omega^c_{\,\,a}
+\frac{2}{3}  \omega^a_{\,\,c} \wedge  \omega^c_{\,\,d} \wedge \omega^d_{\,\,a},
\quad \Omega_{\rm 3Y} = \mathbf{A} \wedge  \mathbf{d} \mathbf{A} + \mathbf{A} \wedge \mathbf{A} \wedge \mathbf{A},
\end{align}
where $\alpha^\prime = M_s^{-2}$, with $M_s$ the string mass scale, which is in general different from the 
four-dimensional Planck mass scale $M_P = 1.22 \times 10^{19}~{\rm GeV} \equiv \sqrt{8\,\pi}\,\kappa^{-1}$. We use standard differential form notation for brevity~\cite{eguchi}. Above, $\mathbf{A}$ is the Yang-Mills gauge field one-form, and $\omega^a_{\,\,b}$ the spin-connection one-form (the Latin indices $a,b,c,d$ are (3+1)-dimensional tangent-space ({\it i.e}. Lorentz-group-SO(1,3)) indices, referring to the Minkowski manifold which is tangent to the space-time manifold at a coordinate point $x$).
The addition of the CS terms to \eq{csterms} leads to a modification of the Bianchi identity (\ref{bianchi}), which can now be written as~\cite{str2}
\begin{align}\label{modbianchi2}
& \varepsilon^{\mu\nu\rho\sigma}\, \mathcal H_{[\nu\rho\sigma\, ;\, \mu]} =  \varepsilon_{abc}^{\;\;\;\;\;\;\mu}\, {\mathcal H}^{abc}_{\;\;\;\;\;\; ;\mu} 
 =  \frac{\alpha^\prime}{32\, \kappa} \, \sqrt{-g}\, \Big(R_{\mu\nu\rho\sigma}\, \widetilde R^{\mu\nu\rho\sigma} -
\mathbf F_{\mu\nu}\, \widetilde{\mathbf F}^{\mu\nu}\Big) \,,
\end{align}
where the right side denotes the mixed anomaly, due to chiral fermions in the theory circulating in the anomalous loop~\cite{alvarez,weinberg},  $g$ denotes the determinant of the metric tensor, $\mathbf{F} = \mathbf{d} \mathbf{A} + \mathbf{A} \wedge  \mathbf{A}$ is the two-form corresponding to the Yang-Mills field strength  (we use form notation for brevity here),  
$R_{\mu\nu\rho\sigma}$ is the Riemann space-time curvature tensor\footnote{Our conventions and definitions used throughout this work are: signature of metric $(+, -,-,- )$, Riemann Curvature tensor
$R^\lambda_{\,\,\,\,\mu \nu \sigma} = \partial_\nu \, \Gamma^\lambda_{\,\,\mu\sigma} + \Gamma^\rho_{\,\, \mu\sigma} \, \Gamma^\lambda_{\,\, \rho\nu} - (\nu \leftrightarrow \sigma)$, Ricci tensor $R_{\mu\nu} = R^\lambda_{\,\,\,\,\mu \lambda \nu}$, and Ricci scalar $R = R_{\mu\nu}g^{\mu\nu}$.} 
and 
\begin{equation}\label{leviC}
\varepsilon_{\mu\nu\rho\sigma} = \sqrt{-g}\,  \epsilon_{\mu\nu\rho\sigma}, \quad \varepsilon^{\mu\nu\rho\sigma} =\frac{{\rm sgn}(g)}{\sqrt{-g}}\,  \epsilon^{\mu\nu\rho\sigma},
\end{equation}
with $\epsilon^{0123} = +1$, {\emph etc.}, are the gravitationally covariant Levi-Civita tensor densities, totally antisymmetric in their indices.
The tilde symbol above the curvature and gauge-field tensors denotes the corresponding dual, defined as
\begin{align}\label{duals}
\widetilde R_{\mu\nu\rho\sigma}= \frac{1}{2} \varepsilon_{\mu\nu\lambda\pi} R_{\,\,\,\,\,\,\,\rho\sigma}^{\lambda\pi}, \quad \widetilde{\mathbf F}_{\mu\nu}=\frac{1}{2} \varepsilon_{\mu\nu\rho\sigma}\, \mathbf F^{\rho\sigma}.
\end{align}
The nonzero quantity on the right hand side  of \eq{modbianchi2} is the ``mixed (gauge and gravitational) quantum anomaly''~\cite{weinberg,alvarez}, which is known to be a total divergence of a topological current $\mathcal K^\mu$ (containing the CS forms in \eq{csterms}). 
In general, the parameters $\alpha^\prime$ and $\kappa^2$ are independent in generic string models~\cite{pol1,pol2}, especially in view of the possibility of large-extra-dimension compactifications, 

Implementing the Bianchi identity 
\eqref{modbianchi2} in a path integral of the low-energy theory, we obtain 
an effective target-space gravitational action
for the ground state of the bosonic sector of superstring theory, of interest to us here. 
To lowest order in the string Regge slope $\alpha^\prime$, the (3+1)-dimensional effective action of the closed-string bosonic sector is then given by
~\cite{str1,str2,pol1,pol2,strom,kaloper}:
\begin{align}\label{sea3}
S^{\rm eff (I)}_{\rm B} =&\; \int d^{4}x\sqrt{-g}\Big[ -\dfrac{1}{2\kappa^{2}}\, R + \frac{1}{2}\, \partial_\mu b \, \partial^\mu b \nonumber \\& +  \sqrt{\frac{2}{3}} \, \frac{\alpha^\prime}{96\, \kappa} \, b(x) \, \Big(R_{\mu\nu\rho\sigma}\, \widetilde R^{\mu\nu\rho\sigma} - \mathbf F_{\mu\nu}\, \widetilde{\mathbf F}^{\mu\nu}\Big) + \dots \Big],
\end{align}
where the dots $\dots$ denote gauge, as well as higher derivative terms appearing in the string effective action, that we ignore for our discussion here.  In this construction, $b(x)$ is the KR axion, which appears initially as a {\it standard pseudoscalar Lagrange multiplier field}, and eventually becomes 
dynamical, after path-integration
of the $\mathcal H_{\mu\nu\rho}$ field~\cite{kaloper}, with a canonically-normalised kinetic term with the {\it correct sign}~\cite{strom} relative to the space-time curvature (Eisntein-Hilbert) terms in \eq{sea3}.

At this point we should remark that, in (3+1) spacetime dimensions, the KR axion field is the dual of the KR field strength $\mathcal H_{\mu\nu\rho}$:
\begin{align}\label{dual2}\partial_\sigma b \propto \varepsilon_{\sigma\mu\nu\rho}\mathcal H^{\mu\nu\rho}\,.
\end{align}
Given that in string theory, up and including $\mathcal O(\alpha^\prime)$ in the effective target-space action, 
$\mathcal H_{\mu\nu\rho}$ plays the r\^ole of spacetime torsion, one may observe that the considerations in our article regarding the model \eqref{sea3}, 
apply also to Einstein-Cartan theories~\cite{Cartan:1938ph}, in which the axionic degrees of freedom are dual (in the above sense \eqref{dual2}) to the totally antisymmetric part of the torsion~\cite{hehl,shapiro,Hammond:2002rm,iorio}. For completion at this stage we also note that, because of \eqref{dual2}, the Euclidean version of the string effective action (to $\mathcal O(\alpha^\prime)$)~\cite{str1,str2,pol1,pol2}: 
\begin{align}\label{sea2}
S_B =-&\; \int d^{4}x\sqrt{-g}\Big( \dfrac{1}{2\kappa^{2}}\, R + \frac{1}{6}\, {\mathcal H}_{\lambda\mu\nu}\, {\mathcal H}^{\lambda\mu\nu} +\dots\Big)\,,
\end{align}
there are ambiguities when we analytically continue \eqref{sea2}, with \eqref{dual2}, back to Minkowski-signature spacetimes. The latter are associated with the order we implement \eqref{dual2} in the path integral~\cite{strom}. Indeed, if we {\it first} use the {\it property} of the Levi-Civita tensor density in four space-time dimensions with Euclidean metric,
\begin{align}\label{propLC}
\varepsilon_{\mu\nu\rho\lambda}^{\rm (E)} \, 
 \varepsilon^{\mu\nu\rho\sigma\, \rm (E)} = + 6 \, \delta_\lambda^{\, \sigma},
 \end{align}
 where $\delta_\lambda^{\, \sigma}$ denotes the Kronecker delta, and {\it then analytically continue} the action \eqref{sea2} to {\it Minkowskiian-signature} space time, we obtain the real (Hermitian) effective gravitational action \eqref{sea3}. However, if we {\it first analytically continues} \eqref{sea2} to a {\it Minkowski-signature space time}, and {\it then use}
the Minkowski-signature version of \eq{propLC}, that is the relation: 
\begin{align}\label{propLCM}
\varepsilon_{\mu\nu\rho\lambda}\, 
\varepsilon^{\mu\nu\rho\sigma} =-6 \delta_\lambda^\sigma~, 
\end{align}
then, we obtain an effective action in which the kinetic terms of the $b$ field have the {\it wrong} sign relative to the space-time-curvature terms in the effective action, and thus the KR axion would behave like a {\it ghost (phantom)} field. As discussed in \cite{R3.14}, one can map such an action into a nonHermitian \cPT version of {\it attractive} CS gravity~\cite{jackiw,Alexander:2009tp}. In our article we stay within the normal axion field case. Our \cPT ~non-Hermitian gravity theory, that represents the phase transition of the Hermitian system at the infrared-region singularities of the RG flow, corresponds to a {\it repulsive} gravitational force at large distances. 

The KR axion is the so-called model-independent string axion~\cite{witten}, associated with the gravitational string multiplet as explained above, which is to be distinguished from other stringy axions arising from compactification. We also observe that, in view of the mixed anomaly appearing in \eqref{modbianchi2}, the KR axion field couples to the gravitational and gauge fields. This latter interaction is $\cP$ and $\cT$ violating; hence, in view of the overall $\cCPT$ invariance of any relativistic, local and unitary (quantum) field theory \eq{sea3}, the interaction is $\cC\cP$ violating (where $\cC$ denotes the Dirac charge-conjugation operator). This is the standard approach to deriving the effective theories of string gravitational axions and their mixing with the axions coming from compactifications~\cite{kaloper,witten}, in which the axions are  pseudoscalar fields. This approach leads to a rich phenomenology~\cite{arvanitaki,marsh,astro2,astro3,astro4,kolbaxion,Manton:2024hyc}.\footnote{We note for completeness that 
an important property of the  Hermitian  
non-Abelian  axion-gauge-Pontryagin-index interactions in \eqref{sea3} is 
periodicity of the effective actions under shifts of the axion field $b(x)$. Indeed, in view of the 
Pontryagin-topological-class interpretation of the gauge CS anomalous action term (in the language of forms for brevity)~\cite{eguchi}:
\begin{align}\label{indexx}
\chi = \frac{1}{16\pi^2} \int d^4x\, {\rm Tr}(\mathbf F \wedge\mathbf F)\in\mathbb Z\,, 
\end{align}
where $\mathbf F$ the non-Abelian gauge field strength two form, we observe that the Hermitian action term involving the CS coupling of the $b$-field:
\begin{align}\label{axion}
S \ni \frac{1}{16\pi^2 \, f_b} \int d^4 x \, b(x)\, {\rm Tr}(\mathbf F \wedge \mathbf F) 
\end{align}
exhibits a periodicity in the $b$ field, $b(x) \to  b(x) + 2\pi\, f_b$, where $f_b$ defines the so-called axion coupling, which in the case of \eq{sea3} is given by 
\begin{align}\label{axionbcoupl}
f_b^{-1}  = \frac{\pi^2}{2} \, \sqrt{\frac{1}{6}}\, \frac{\alpha^\prime}{\kappa}\,.
\end{align}
This implies the breaking of the shift-symmetry $ b(x) \to b(x) + {\rm constant}$ (due to the total derivative nature of the anomalous Pointryagin terms in \eqref{sea3}, 
by instanton effects in the non-Abelian gauge theories for which the index $\chi\ne0$ \eq{indexx}, and leads to the well-known nonperturbative axion potentials $V(b)= \Lambda^4_{\rm inst}\big[1-\cos(b/f_b)\big]$, and axion masses of order $m_b\sim\Lambda_{\rm inst}^2/f_b$, where $\Lambda_{\rm inst}$ is the scale of the non-Abelian gauge group instanton effects. \cbl}

For our purposes we reduce the gauge sector to only Abelian ($U(1)$ electrrodynamics) terms. In what follows we  ignore, for the moment, the gravitational dynamics placing the rest of the terms in \eqref{sea3} in flat target spacetimes. 
The \cm resulting \cbl  part of the action \eqref{sea3} then leads to an axion electrodynamics model discussed in Sec.~\ref{s3}, which is quantised using path-integral and RG analysis nonperturbatively for the axion coupling~\cite{Eichhorn:2012uv}. This reveals a crucial singular behavior (in infrared regions of the energy scale) in the phase space, which we bypass by means of replacing (near an appropriate infrared fixed point) the Hermitian model by a suitable non-Hermitian but \cPT-symmetric model, obtained via the replacement of the axion coupling $f_b$ \eqref{axionbcoupl} by a purely imaginary one:\footnote{The reader should notice that the resulting nonHermitian field theory no longer exhibits a periodicity in the axion field $b(x)$ unlike the Hermitian case. Axion masses in such a case may be generated dynamically, e.g. through their Yukawa interactions with appropriate chiral fermions~\cite{R3.13,Alexandre:2020tba,R3.14,R3a,R3c}, including heavy right-handed neutrinos~\cite{Mavromatos:2023bdx}, that may appear in extensions of the Standard Model.} 
\begin{align}\label{replfb}
f_b \quad \to \quad 
if_b\,.
\end{align}
The replacement \eqref{replfb} signifies a new phase. We then use such flat-spacetime arguments as support for our conjecture on the emergence of a {\it repulsive phase} of gravity at {\it large scales} (infrared), once we embed the model in a dynamical curved spacetime. 

\cb In the next section, we revisit a non-perturbsative (functional) renormalization-group (RG) analysis of the flat-space axion electrodynamics, given in 
\cite{Eichhorn:2012uv}. This model constitutes (the flat-spacetime) part of the effective string-inspired Chern-Simons gravity model, and its RG structure will be used to argue  the existence of a \cPT-symmetric phase of gravity, when the model is embedded in curved spacetimes.  \cm The non-perturbative RG flow given in \cite{Eichhorn:2012uv} is an important technical input to show the singularity in the RG flow of the Hermitian theory and also to discuss the flow in the associated \cPT-symmetric theory.  Hence we first review some of the analysis in \cite{Eichhorn:2012uv}  to make our argument clear. Our interpretation in terms of \cPT~phases given below is completely new, and constitutes the main purpose of the current manuscript.\cbl  

\section{A reinterpretation of the Non-perturbative Renormalization Group Analysis of Axion Electrodynamics} \label{s3}

We 
commence our discussion by 
first reviewing a non-perturbative RG analysis \cite{Eichhorn:2012uv} of the Euclidean Lagrangian $\mathcal{L}_E$ related to \eqref{axion} in the massless ($m \to 0$) limit:
\be 
\label{eq:AxEl}
\mathcal{L}_{E}=\frac{1}{4} F_{\mu \nu }F_{\mu \nu }+\frac{1}{2} \partial_{\mu } a\partial_{\mu } a+\frac{1}{2} m^{2}a^{2}+\frac{1}{4} i\mathfrak{g}\, a\, F_{\mu \nu }\tilde{F}_{\mu \nu } \,,
\ee
with gauge fixing $\nabla .{\bf A}=0\  \  \  \rm {and}  \    \  \  A_{0}=0$. The mass dimension of $\mathfrak{g}$ is $-1$ and so the theory is not perturbatively renormalisable.\footnote{At this stage the reader should recall, for future use, that in the context of our string theory \eqref{sea3}, the axion field $a(x)$ corresponds to the KR massless axion $b(x)$, and the coupling $\mathfrak{g}$ is given by $\mathfrak{g} =  \sqrt{\frac{2}{3}} \, \frac{\alpha^\prime}{24\, \kappa} = \frac{1}{4\pi^2\, f_b}$, where $f_p$ is defined in \eqref{axionbcoupl}.}

This theory, however, can be treated using the functional renormalisation group (FRG) and the Wetterich equation \cite{wetterich1993exact,Berges:2000ew,Wipf:2013vp}.
\be
\label{eq:Wetterich}
\partial_{t} \Gamma_{k} =\tfrac{1}{2} Tr[ \partial_{t} R_{k}( \Gamma^{( 2)  }_{k} +R_{k})^{-1}]  
\ee
where $\frac{\partial }{\partial t} \equiv k\frac{d}{dk} $, with $k$ a running RG scale.
$\Gamma_{k}$ is the effective average action. $\Gamma_{k\rightarrow \Lambda } =S_{\Lambda }$ is the bare action and $\Lambda$ is the UV cut-off.~$\Gamma_{k\rightarrow 0} =\Gamma $ is the full quantum effective action. The Wetterich equation is an integro-differential nonlinear equation and cannot be solved exactly. Progress in obtaining solutions requires an ansatz for $\Gamma_{k}$.
A standard ansatz restricts $\Gamma_{k}$ to
\be
\label{eq:Ansatz}
\Gamma_{k} =\int d^{4}x[ \tfrac{Z_{F}}{4} ( F_{{}_{\mu \nu }}( x)  )^{2}  +\tfrac{Z_{a}}{2} \left( \partial_{\mu } a\right)^{2}  +\tfrac{m^{2}_{k}}{2} a\left( x\right)^{2}  +\tfrac{i\mathfrak{g}_{k}}{4} a\left( x\right)  F_{\mu \nu }\left( x\right)  \tilde{F}_{\mu \nu } \left( x\right) ]  +Z_{F}\int d^{4}x\,L_{GF}
\ee
where $L_{GF}$ denotes a gauge fixing piece to the Lagrangian, whose details can be found in  \cite{Eichhorn:2012uv}. The beta functions calculated using this scheme are not universal but are scheme dependent \cite{Eichhorn:2012uv}. This is a feature of \emph{non-renormalisability} in the perturbative sense.

From the above ansatz it is shown in ~\cite{Eichhorn:2012uv} that 
\be
\tfrac{\partial }{\partial t} m^{2}_{k}=\tfrac{\partial }{\partial t} \mathfrak{g}_{k}=0.
\ee
The physical observables are expressed in terms of dimensionful quantities:
\be
\label{eq:PhyObs}
m^{2}_{R}=\tfrac{m^{2}_{k}}{Z_{a}} ,\  \  \  \  \  \mathfrak{g}^{2}_{R}=\tfrac{\mathfrak{g}^{2}_{k}}{Z^{2}_{F}Z_{a}} 
\ee
where 
\begin{align}
\partial_{t} Z_{a}&=\tfrac{1}{\Omega } \left[ \tfrac{\partial^{2} }{\partial q^{2}} \int d^{4}p\tfrac{\delta^{2} }{\delta a\left( p\right)  \delta a\left( -q\right)  } \partial_{t} \Gamma_{k} \right]_{a,A,q\rightarrow 0}, \\
\partial_{t} Z_{F}&=\tfrac{1}{\Omega } \left[ \tfrac{4}{3} \tfrac{\partial^{2} }{\partial q^{2}} \int d^{4}p~n_{\kappa }n_{\lambda }\tfrac{\delta^{2} }{\delta A_{\kappa }\left( p\right)  \delta A_{\lambda }\left( -q\right)  } \partial_{t} \Gamma_{k} \right]_{a,A,q\rightarrow 0}  
\end{align}
This is a standard procedure in applying the functional renormalisation group. Associated dimensionless quantities are
\begin{align}\label{dimensionless}
m^{2} \equiv \tfrac{m^{2}_{R}}{k^{2}}\,, \quad   g^{2} \equiv {\mathfrak{g}^{2}_{R}}k^{2}\,.
\end{align}
We define the field anomalous dimensions as
\be \label{eq:AnomDim}
\gamma_{a} =-\partial_{t} \rm {log}Z_{a},\  \  \  \  \  \gamma_{F} =-\partial_{t} \rm {log}Z_{F}.\  \  
\ee
This leads to the following nonperturbative beta functions, which determine the running of the coupling constants:
\begin{align}
\label{eq:runningCouplings}
 \partial_{t} g^{2}&=\left( 2+2\gamma_{F} +\gamma_{a} \right)  g^{2},\\  
 \partial_{t} m^{2}&=\left( \gamma_{a} -2\right)  m^{2}.
\end{align}
where 
\begin{align}
\label{eq:ExplicitAnomDim}
 \gamma_{a} &=\frac{g^{2}}{6\left( 4\pi \right)^{2}  } \left( 2-\frac{\gamma_{F} }{4} \right),\\ 
\gamma_{F} &=\frac{g^{2}}{6( 4\pi )^{2}  }\left ( \frac{( 2-\tfrac{\gamma_{a} }{4} )  }{( 1+m^{2})^{2}  } +\frac{( 2-\tfrac{\gamma_{F} }{4} )  }{1+m^{2}} \right ) 
\end{align}  

These equations, derived in \cite{Eichhorn:2012uv}, are the starting point of our analysis. They  have not been carefully analysed for the full range of $m$ and $g$. However, the analysis of various limiting cases is still relevant for us in conjunction with relations of the form given in  \eqref{irG}; these will be discussed in the next subsection. 

\subsection{Behavior of renormalisation group flows in limiting cases}

For $m\gg 1$ and $g\ll 1$ (weak coupling), 
\be
\partial_{t} g^{2}=\left( 2+\frac{g^{2}}{3\left( 4\pi \right)^{2}  } \right)  g^{2}
\ee
which has a solution $g^{2}_{R}\left( k\right)  =\frac{g^{2}_{R}\left( \Lambda \right)  }{1+\frac{1}{6\left( 4\pi \right)^{2}  } \left( \Lambda^{2} -k^{2}\right)  g^{2}_{R}\left( \Lambda \right)  } $.
Also $m^{2}_{R}\left( k\right)  =\frac{m^{2}_{R}\left( \Lambda \right)  }{1+\frac{1}{6\left( 4\pi \right)^{2}  } \left( \Lambda^{2} -k^{2}\right)  g^{2}_{R}\left( \Lambda \right)  } $.
    
     Let us consider the massless theory, which is appropriate for the flat spacetime and Abelian gauge sector limit of our string-inspired gravitational axion theory \eqref{sea3}, discussed in the previous section. In this case, from  \eqref{eq:runningCouplings} we deduce that $m=0$ is an infrared fixed point. For $m=0$  and $ g\ll 1 $
     \be
     \label{eq:strongcoupling}
     \partial_{t} g^{2}=\left( 2+\frac{5g^{2}}{3\left( 4\pi \right)^{2}  } \right)  g^{2}.
     \ee
This leads to
\be
\mathfrak{g}^{2}_{R}\left( k\right)  =\frac{\mathfrak{g}^{2}_{R}\left( \Lambda \right)  }{1+\frac{5}{6\left( 4\pi \right)^{2}  } \left( \Lambda^{2} -k^{2}\right)  \mathfrak{g}^{2}_{R}\left( \Lambda \right)  }, 
\ee
a  result  similar to that found for the case $m\gg 1$, examined above. Both these results do not indicate any interesting non-analyticity in the flow to the infrared. We move away from these limits to see whether there is any nonanalytic structure.

For arbitrary values of $g$ we are able to solve 
      \eqref{eq:ExplicitAnomDim} for $\gamma_{a}$ and $ \gamma_{F} $  (as discussed in detail in~\cite{Eichhorn:2012uv}, where we refer the interested reader for further study)
      \begin{align}
      \partial_{t} g^{2}&=\beta_{g^{2}} =2g^{2}\frac{13g^{4}-384\pi^{2} g^{2}\left( 21+17m^{2}+4m^{4}\right)  -147456\pi^{4} \left( 1+m^{2}\right)^{2}  }{g^{4}-384\pi^{2} g^{2}\left( 1+m^{2}\right)  -147456\pi^{4} \left( 1+m^{2}\right)^{2}  }\\ \partial_{t} m^{2}&=\beta_{m^{2}} =6m^{2}\frac{-g^{4}+128\pi^{2} g^{2}\left( 3+7m^{2}+4m^{4}\right)  -49152\pi^{4} \left( 1+m^{2}\right)^{2}  }{-g^{4}+384\pi^{2} g^{2}\left( 1+m^{2}\right)  +147456\pi^{4} \left( 1+m^{2}\right)^{2}  }     
      \end{align}
      This form of the beta functions   is manifestly nonperturbative. Our application of the solution of these equations to Chern-Simons gravity and an emergence of $\cPT$ symmetry is new. For $m=0$ the expressions simplify and we are able to solve these equations, which is crucial for our analysis.
    
  \subsection{Massless case: infrared singularities in the non-perturbative renormalisation group flows and a \cPT symmetric phase}
       
\begin{figure}[ht]
  \centering
  \includegraphics[width=0.8\textwidth]{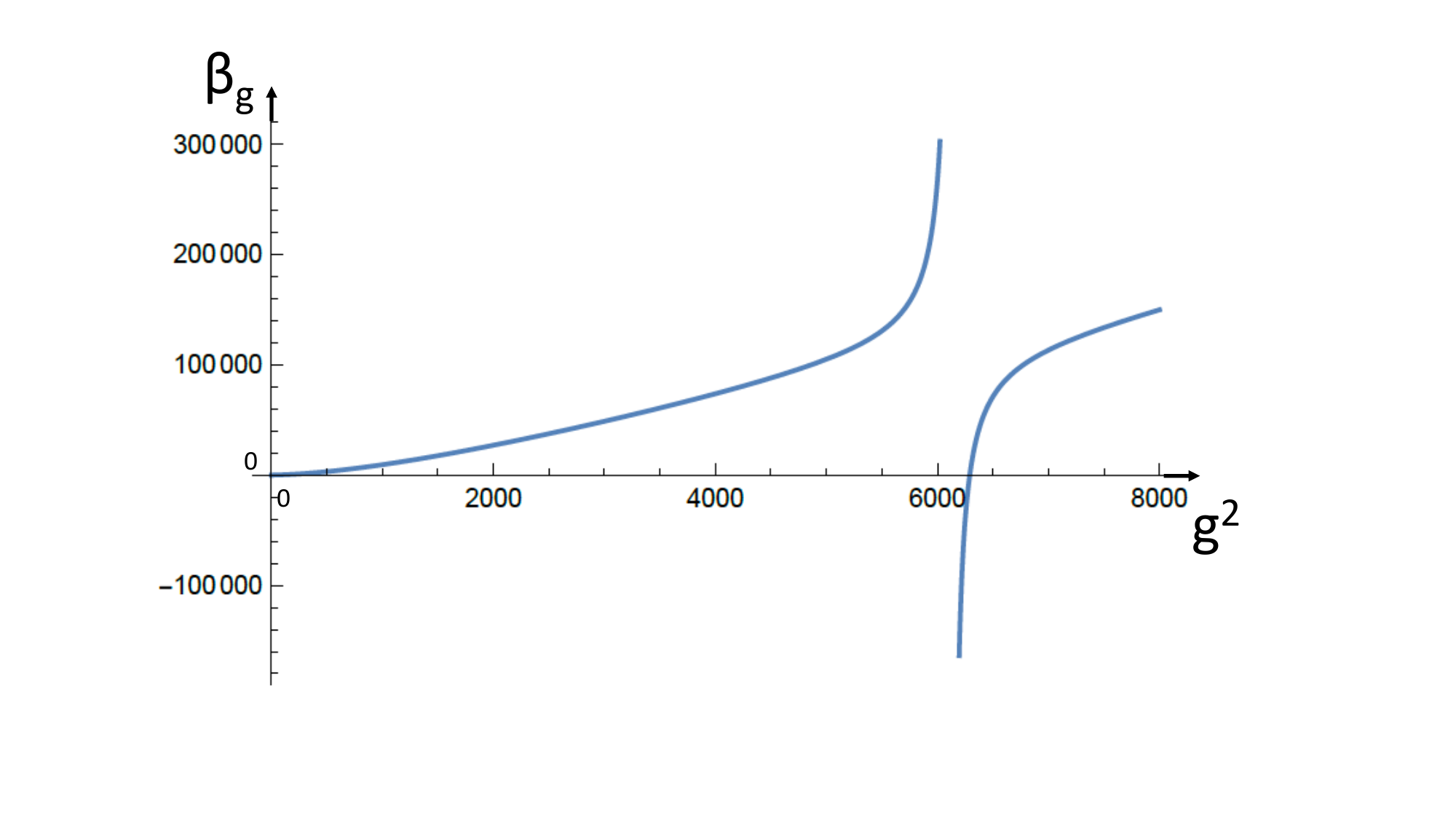}  
  \caption{The beta function $\beta_g \equiv \partial_t g^2$ of the Hermitian theory, given in Eq.~\eqref{eq:Arbg}, exhibiting the proximity of a singular behavior at a finite running coupling \eqref{gvalaxqed}, as well as a fixed point behavior ($\beta_g (g)\big|_{g=g_\star}=0$), characterised by the existence of a trivial fixed point at zero coupling, and a non-trivial one at strong coupling, $g_\star^2 \gg 1$. A similar figure appears in \cite{Eichhorn:2012uv}.}
  \label{fig:betagplot}
\end{figure}

      We consider  the massless case, $m=0$, but \emph{arbitrary} $g$, studied in~\cite{Eichhorn:2012uv}, which corresponds to the flat spacetime and Abelian gauge limit of the string effective action \eqref{sea3}. 
       The corresponding beta function reads:\footnote{This equation can be solved using initial conditions determined, e.g. from astrophysical data for axion physics~\cite{PhysRevLett.125.221301}, if we accept the r\^ole of the KR axion as a standard axion. This might not be the case within the most general framework of string theory, so other constraints might be applicable. For our purposes such constants are not relevant, as we are dealing with ratios of couplings for appropriate points in the RG flow.}
       \be
       \label{eq:Arbg}
       \partial_{t} g^{2}=2g^{2}\frac{13g^{4}\  -\  8064\  \pi^{2} g^{2}-147456\pi^{4} }{g^{4}-384\pi^{2} g^{2}-147456\pi^{4} } 
       \ee
The beta function is plotted in Fig.~\ref{fig:betagplot} as a function of $g^2$. The values of $g$ for which the denominator vanishes (thus leading to a singular behavior of the beta function) have the property that they are either real or purely imaginary and come in  pairs: 
\begin{align}\label{betasing} {\rm Singularity~of~\beta_{g}~occurs~for}~g_{\rm sing}: (\pm \ 48.3728  i,\  \  \pm 78.2689)\,. 
\end{align}     
This is significant and leads to a new \emph{$\cPT$ symmetric interpretation}, which crucial for our analysis. Hence, presumably, the beta function of a non-Hermitan theory would also show such singular behavior, as we shall show later. Of interest in what follows will be the singularity of the Hermitian theory, occurring at ({\it cf.} \eqref{betasing}):
\be\label{gvalaxqed}
g_{\rm sing}^2 = (\mathfrak{g}_{\rm sing\,R} (k) \, k )^2 \,{\Big| _{k=k_{\rm sing}}}= 6126.0207\,, 
\ee
which corresponds to a RG momentum scale $k=k_{\rm sing} > 0$.

Next we look at the fixed points of the beta function of the \emph{Hermitian} theory, {\it i.e.} the values of the coupling at which the beta function vanishes. Notably, \eqref{eq:Arbg} has a trivial {\it IR attractive fixed point} at zero coupling $g=g_\star^{\rm IR}=0$, as well as two real and two purely imaginary \emph{non-trivial fixed points}, but no fixed points with both real and imaginary parts. Explicitly the non-trivial fixed points denoted by $g^\star$ are:  
\begin{align}\label{fphermit}
g^\star = (\pm 13.2389\  i,\  \  \pm 79.3174)\,.
\end{align}
For the Hermitian theory therefore, the nontrivial IR stable fixed point occurs at 
\begin{align}\label{fphermit2}
{g^\star}^2 = 6291.2499\,.
\end{align}
This is an IR fixed point as seen from the form of the beta function in the neighbourhood of this fixed point (passing from negative to positive, as $g^2$ increases). 
Thus, the Hermitian theory has an {\it IR attractive fixed point} at zero coupling $g=g_\star^{\rm IR}=0$, followed  (as $g^2$ increases) by a singular behavior, which occurs at a RG momentum scale $k=k_{\rm sing} > 0$, corresponding to the value \eqref{gvalaxqed} of $g$.

We now examine the scale $k=k_{\rm sing}$ corresponding to the singularity \eqref{gvalaxqed} of the RG $\beta$ function near the non-trivial IR fixed point (see fig.~\ref{fig:betagplot}). We estimate the value of  $k_{\rm sing}$ by solving the RG  equation \eqref{eq:Arbg} to obtain $g=g(t)$. On integrating \eqref{eq:Arbg} from the nontrivial IR fixed point \eqref{fphermit2}, corresponding to a RG scale $k_0 \equiv m_{\rm IR}$, to a 
point corresponding to an RG scale $k$, 
we obtain\footnote{In our analysis we denote $\sqrt{g^2(k)}$ by $|g(k)|$. } 
\begin{align}\label{gsolherm} 
|g(k)| \, (2281 + 13 \,g^2(k))^{-0.46}\, |27.6 - 4.4 \cdot 10^{-3}\, g^2(k)|^{0.0013}   = a e^{t-t_0} = a \frac{k}{k_0}\,.
\end{align}
where ({\it cf.} \eqref{fphermit2}) 
$a\equiv |g^\star | \, (2281 + 13 \,g^{\star 2})^{-0.46}\, |27.6 - 4.4 \cdot 10^{-3}\, g^{\star 2}|^{0.0013} \simeq 0.4$,  and thus, the RG running scale $k_{sing}$, corresponding to the singularity \eqref{gvalaxqed}, is given by:
\begin{align}\label{ksing}
k_{sing} = 1.01 \, k_{0}\,. 
\end{align}
The solution in \eqref{gsolherm} is just the principal branch of a function $g(k)$ with an \emph{infinite Riemann sheet} structure and so  is extremely complicated. The singularity structure and non-analyticity indicate that the underlying approximations are inadequate, and so the field theory may exist in a different phase (as occurs in an anharmonic oscillator in quantum mechanics where a much more complete analysis is possible).
From \eqref{gsolherm}  we deduce at $k=k_{sing}$ immediately a very similar relation for $g_{sing}$. This result is consistent since in this range of $g^2$ the beta function is negative; hence in the RG flow from the region of the singularity \eqref{gvalaxqed} to the non-trivial IR fixed point \eqref{fphermit2}, an increasing running $g^2(k)$ corresponds to a decreasing RG running scale $k$.
Formally, we can take $k_0 \to 0$, indicating that the 
singularity of the Hermitian theory lies very near the non-trivial IR fixed point. 
Physically $k_0$ could be taken to be the energy scale corresponding to the inverse size of the Universe, for our purposes. 

For $g^2 > g_{\rm sing}^2$ the beta function has (very large) negative values until the (nearby) non-trivial IR fixed point  ${g_\star}^2 > g_{\rm sing}^2$ is reached, at which the beta function vanishes. After this second IR fixed point the beta function is positive, increasing monotonically (in an unbounded way) with increasing $g$. 
So as $g^2 \to \infty$ the beta function tends  to infinity. Indeed in the region of very large positive $g^2$, the beta function  equation \eqref{eq:Arbg} can be approximated by
\begin{align}\label{largeg}
       \partial_{t} g^{2} \simeq 26\, g^{2}\,, \qquad g^2 \gg {g^{\star}}^2 = 6291.2499\,.
\end{align} 
and so as $ g \to \infty$, the above equation is easily integrated to give analytically the scaling~\cite{Eichhorn:2012uv}:
\begin{align}\label{ginfty}
    \frac{{\rm g^2(k_\infty)}}{g^2(k)} \sim e^{26(t_\infty - t)} = \Big(\frac{k_\infty}{k}\Big)^{26}\,, \quad g^2(k) \gg 1, \quad k < k_\infty\,,
\end{align}
where $k_\infty \sim e^{t_\infty}$ is a large RG scale at which $ g( k_\infty) $ is very large; so the coupling diverges highly non-linearly with the RG scale (large anomalous dumension). 
We next remark that, if we adopt the point of view of effective field theory, that no RG scale can exceed the UV cutoff $\Lambda$, then in Eq.~\eqref{ginfty} 
 we must identify $k_\infty $ with $\Lambda$. In  our string-inspired model, we may identify $\Lambda=M_s$, where $M_s$ is the string scale, which is assumed not renormalised, {\it i.e} it does not run with the scale $k$. In the pertinent part of the diagram of fig.~\ref{fig:betagplot}, then, the RG scale $k$ flows from $k_0 \sim 0$ to $M_s$, in agreement with the positive signature of the corresponding beta function.

The plot in Fig.~\ref{fig:betagplot} is problematic below the nontrivial IR fixed point, due to the singular behavior of the beta function which lies close to the IR fixed point. We would like to connect the interacting theory to the free theory at $g=0$ smoothly via an appropriate RG flow from an UV to IR fixed point. This appears not to be possible if we stay exclusively within the Hermitian theory, as a result of the beta-function singularity at \eqref{gvalaxqed}. 

 In the following we treat the singular behavior of the Hermitian beta function at $k=k_{\rm sing}$ as a {\it phase transition} in the system by considering the corresponding non-Hermitian theory obtained via\footnote{We note for completion that, the non-perturbative RG study of this nonHermitian axion electrodynamics model has been discussed in the Appendix of \cite{Eichhorn:2012uv}, without, however, making any attempt to give that theory any physical significance, or 
 the connection with the \cPT phase of gravity we are conjecturing in the present article.}~$g\to ig$. Notably, this is {\it not} an analytic continuation. This is due to the different RG behavior of this theory. We therefore replace the part of the diagram in Fig.~\ref{fig:betagplot} to the left of (and including) the singularity until the trivial IR fixed point by the corresponding flow diagram pertaining to the non-Hermitian $\cPT$-symmetric theory.
\begin{figure}[ht]
\centering
\includegraphics[width=0.5\textwidth]{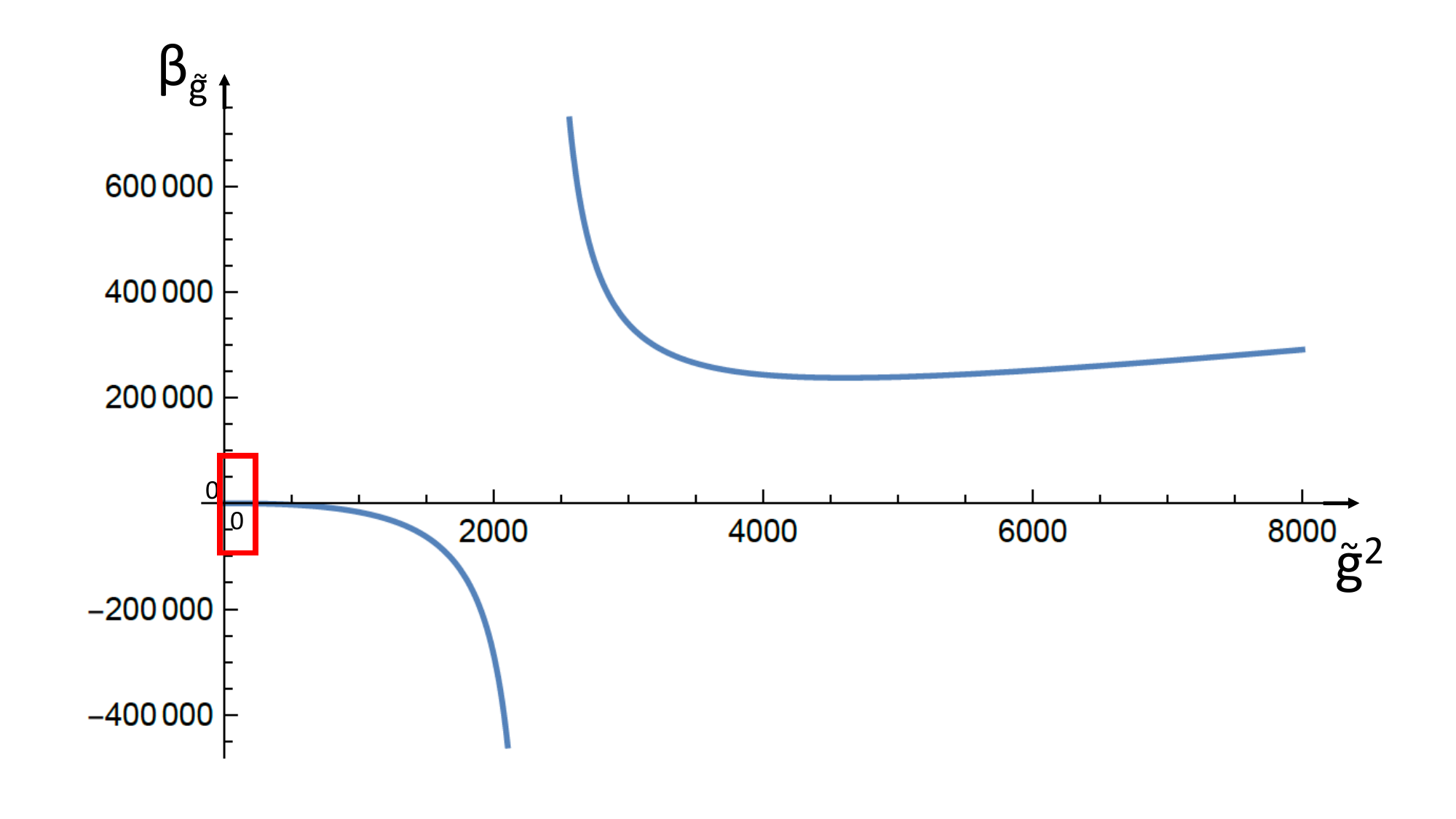} 
\vfill
\includegraphics[width=0.5\textwidth]{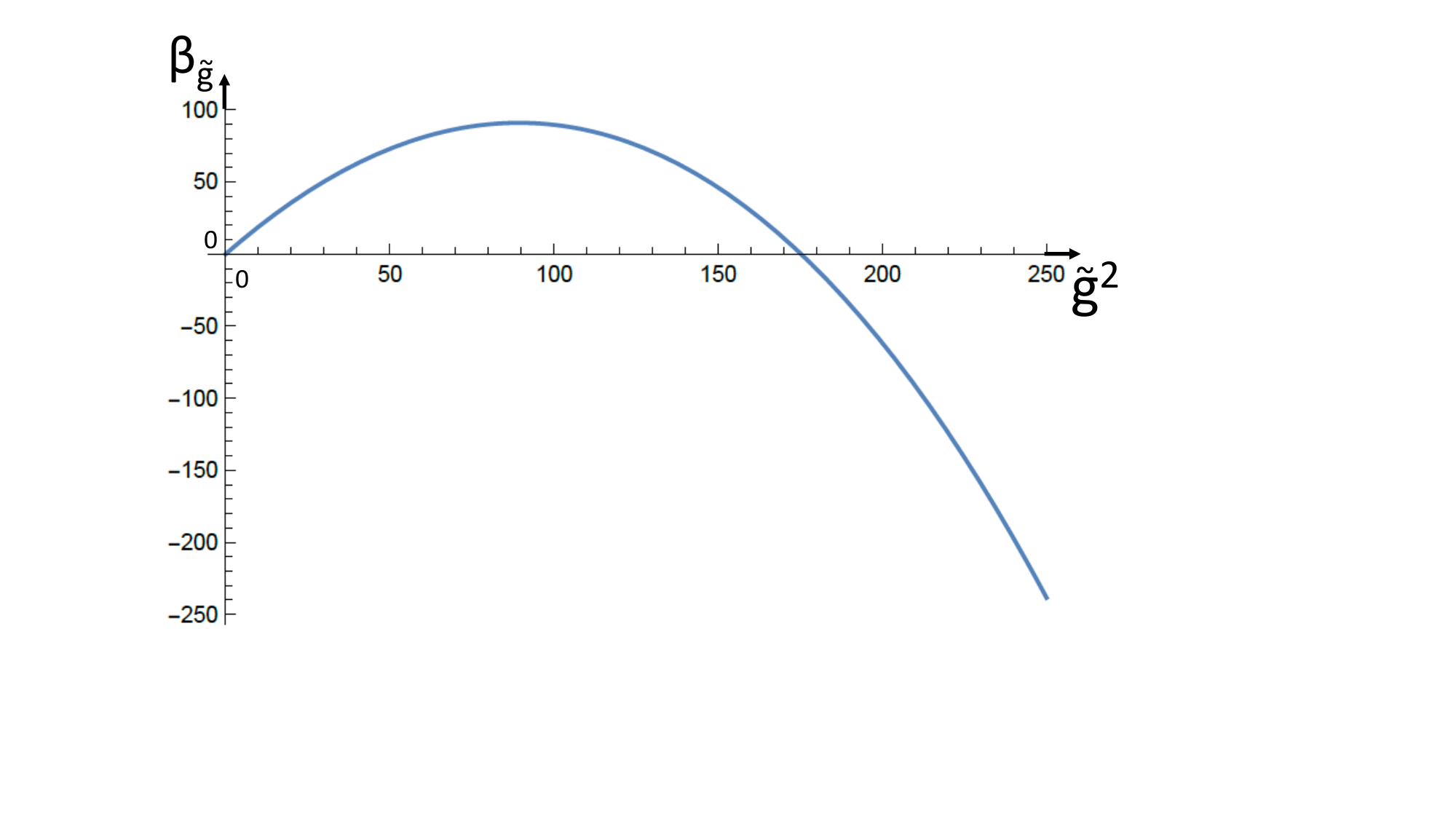}
\vfill
\includegraphics[width=0.5\textwidth]{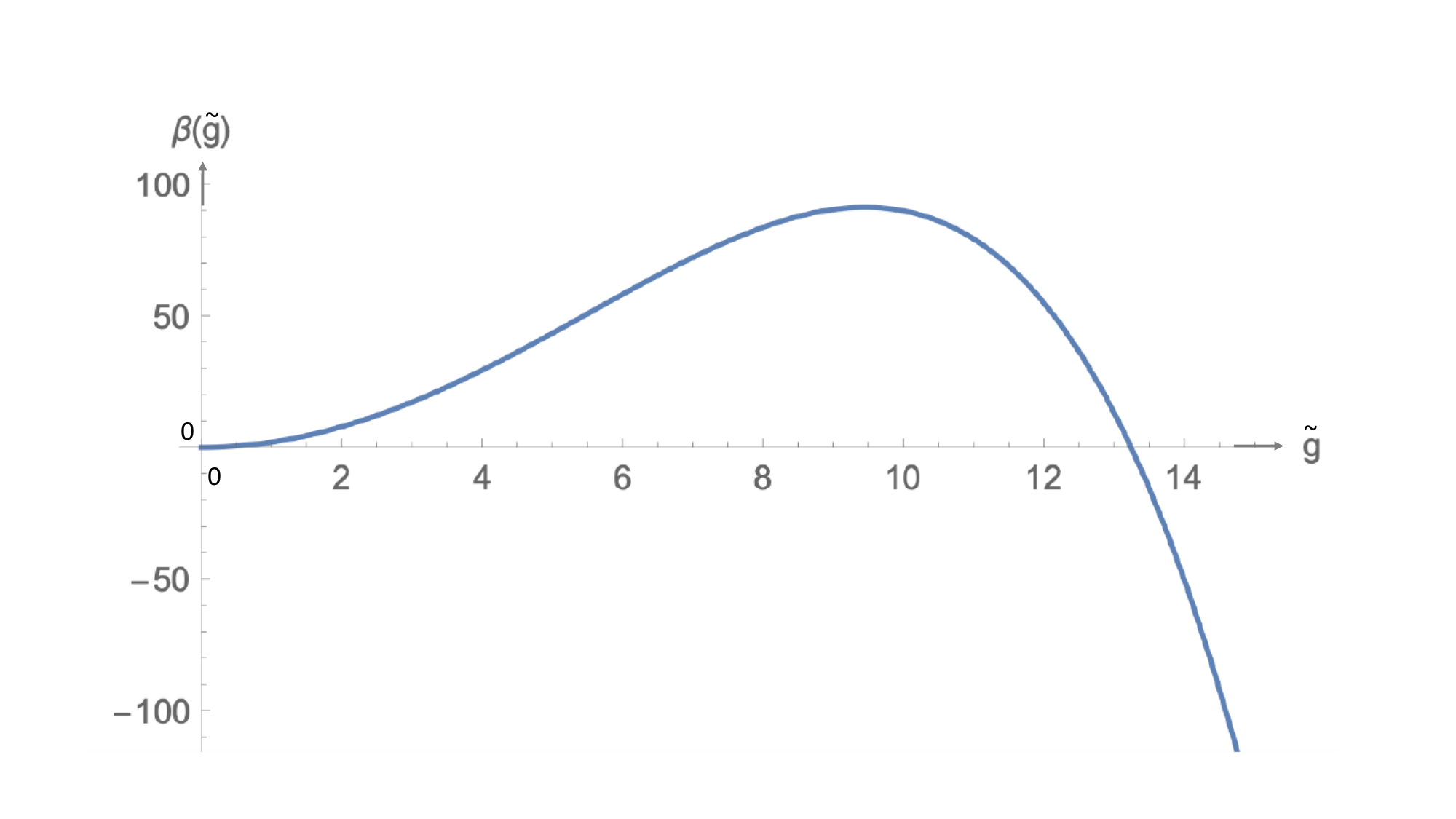}
 \caption{Nonperturbative RG beta function for non-Hermitian case as a function of $\tilde g^2$, given in Eq.~\eqref{eq:ArbgPT} is qualitatively different from the Hermitian beta function. The area included in the red rectangle includes a nontrivial fixed point (apart from the trivial one at zero coupling), which is indicated explicitly in the middle panel. The lower panel is added for completeness, and shows the behavior of the beta function as a function of $\tilde g >0$, in the neighborhood of its fixed points. The middle figure is similar to that in \cite{Eichhorn:2012uv}.}
\label{fig:Beta fn}
\end{figure}

The non-Hermitian theory can be studied explicitly by writing $g=i\tilde{g} $. In this case, we obtain~\cite{Eichhorn:2012uv}
\be\label{eq:ArbgPT}
\partial_{t} \tilde{g}^{2}=\beta_{\tilde{g}}=2\tilde{g}^{2}\frac{13\tilde{g}^{4}\  +\  8064\  \pi^{2} \tilde{g}^{2}-147456\pi^{4} }{\tilde{g}^{4}+384\pi^{2} \tilde{g}^{2}-147456\pi^{4} }\,. 
\ee
The beta function  exhibits a singularity at 
$g^2_{\rm sing~NH} = 2339.93$ and nontrivial fixed points at $\tilde{g}^*_{\rm NH}=\pm 13.2456~~\Rightarrow~~\tilde{g}_{\rm NH}^{*2}= 175.4459$.
We  give plots of $\beta_{\tilde{g}}$ as a function of $\tilde g^2$ (and $\tilde g$)  in Fig.~\ref{fig:Beta fn}. The nontrivial fixed point, which is UV in nature due to the form of the beta function (turning from positive to negative, as $g^2$ increases, see fig.~\ref{fig:Beta fn}),  
is asymptotically free for this Abelian case. Thus, the \emph{RG scale at the non-trivial fixed point can be identified with the UV cutoff $k^\star=\Lambda$}. 
The trivial fixed point is an infrared (IR) attractor. 

As in the Hermitian case, by integrating the beta function from the UV fixed point (corresponding to $k=k_{\rm NH}^\star =\Lambda)$ to the scale $k_{\rm NH \, sing}$, corresponding to the singularity of the beta function ({\it cf.} fig.~\ref{fig:Beta fn}), we obtain:
\begin{align}\label{nhbsol} 
&{|\tilde g_{\rm sing\, NH}| \, |2281 - 13 \,\tilde g_{\rm sing\, NH}^2|^{-0.46}\, (27.6 + 4.4 \cdot 10^{-3}\, \tilde g_{\rm sing\, NH}^2)^{0.0013}} \nonumber \\&= a_{NH} e^{t-t^\star}  = a_{NH} k_{\rm NH \, sing}/k_{\rm NH}^{\star} \,,
\end{align}
(where $a_{NH}\equiv |\tilde g_{\rm NH}^\star | \, |2281 - 13 \,\tilde g_{\rm NH}^{\star 2}|^{-0.46}\, (27.6 + 4.4 \cdot 10^{-3}\, \tilde g_{\rm NH}^{\star 2})^{0.0013}$) which implies that the singularity of the beta function of the non-Hermitian theory occurs at:
\begin{align}\label{ksingnH}
k_{\rm NH \, sing} \simeq 0.34 \Lambda 
\end{align}
This is a consistent result. The beta function is negative for $g^2 > {g^\star_{\rm NH}}^2$, which implies that  $g^2$ should increase for decreasing $k$, in agreement with the above result. But the singularity here occurs in the UV regime.

The singularity of the beta function of the Hermitian theory, which corresponds to a RG scale $k_{\rm sing}$, \eqref{ksing}, 
is assumed in our approach to correspond to a phase transition of the system, which for $k \lesssim k_{\rm sing}$ is described by the non-Hermitian theory. The reader should bear in mind that the scale $k_{\rm sing} \ll k_{\rm NHsing}=0.34\Lambda =0.34 k_{\rm NH}^\star$. Since the beta function of the non-Hermitian theory is positive, this implies that in the region 
of the RG scales $k \le k_{\rm sing}$, $g_{\rm NH}(k) < g_{\rm NH\, sing}$, and thus the corresponding portion of the nonHermitian theory connects \emph{smoothly} (via a RG flow) to the trivial fixed point $g=0$ with $g \ne 0$.

\section{Interpretation in terms of a Repulsive Gravity phase}\label{sec:repgrav}
We now make inferences for the scale dependence of Newton's constant from the above results. Our findings, using axion electrodynamics, are embedded into the discussion of string-inspired gravity with effective action given by \eqref{sea3}.
The analogue of the dimensionless running (renormalised) axion coupling (defined in 
\eqref{eq:AxEl}) is given by: 
\begin{align}\label{axionstringcoupl}
g = \mathfrak{g}_R k =  \frac{1}{24}\, \sqrt{\frac{2}{3}}  \, \Big(\frac{M_{\rm Pl, R}(k)}{M_{s,R}^2(k)}\Big)\, k\,, 
\end{align}
where 
the index $R$ denotes a renormalised quantity, and $k$ is a ``running'' (RG) momentum scale. Our assumption is that when we embed this theory into a \emph{quantum} string-inspired gravity version, the effective \emph{low-energy} theory will also be characterized by a renormalized combination of scales $\Big(\frac{M_{\rm Pl}(k)}{M_s(k)}\Big)_R$, running with the scale $k$. Since the string mass scale $M_s$ is associated with the UV momentum cutoff of the effective low-energy theory, it is reasonable to assume that $M_s$ is not renormalized, {\it i.e.} it does not run with the RG scale $k$.
The effective four-dimensional $M_{\rm Pl}$ is renormalized and runs since it is defined as the effective gravitational scale of the (3+1)-dimensional effective theory.\footnote{In conventional string theory $M_{\rm Pl}$ is proportional to the string-compactification volume and runs with $k$.}
Hence the square of the dimensionless  running (renormalised ($R$))) axion coupling  \eqref{axionstringcoupl} is $g^2=\frac{1}{864} \frac{1}{\widetilde{\rm G}_{\rm N}} \, \frac{k^2}{M_s^2}\,$ 
where the dimensionless quantity $\widetilde{\rm G}_{\rm N} \equiv {\rm G}_{\rm N} \, M_s^2 $, with $\rm G_N$ the (3+1)-dimensional gravitational (Netwon's) coupling,  

In the case of the effective string-inspired low energy theory, we  define $t \equiv \log\Big(\frac{k}{M_s}\Big)$
which implies that the ultraviolet (UV) region is described by the limit $k \to M_s$, {\it i.e.} $t_{\rm UV} \to 0$.
By contrast, the infrared (IR) region corresponds to $k \to 0$ ( or  $k \to k_0 = m_{\rm IR} \ll M_s$, with $m_{\rm IR}$ an infrared mass cutoff).\footnote{When we discussed nonperturbative  renormalisation of the axion electrodynamics field theory \eqref{eq:AxEl}, we gave an alternative definition of the running RG parameter $t$, in terms of an IR scale $k_0$: $t=\log(k/k_0)$, with $k_0 = m_{\rm IR}$, which eventually can be taken to zero, unless there are IR divergencies.  In such a case the UV region corresponds to large $t \gg 1$ (formally $t \to \infty$), whilst the IR regime is described by $t \to 0$. The connection with our case  is to subtract from $t$ the (formally infinite) constant $\log(k_0/M_s)$.} 
The RG running of $\widetilde{\rm G}_{\rm N}(k)$ with $k$, or equivalently with $t$,  is obtained by differentiating both sides of \eqref{axionstringcoupl}
with respect to $t$. This leads to:
\begin{align}\label{gGrun}
\frac{d}{dt}\log\Big(g^2\,\widetilde{\rm  G}_{\rm N} \Big)= 2 \quad \Rightarrow \quad g^2 \, \widetilde{\rm  G}_{\rm N} = \mathfrak{D} \exp\Big(2t\Big)\,,
\end{align} 
where $\mathfrak{D}$ is a positive integration constant to be fixed by the boundary conditions of the RG flow. Considering the equation in the UV regime, $t\to 0$, we have 
$\mathfrak{D} = g^2(t=0)\, \widetilde{\rm  G}_{\rm N}(t=0)$.  For finite $k=M_s < \infty$, the value of $g(k=M_s)\approx g(t=0)$ is large but finite, while the dimensionless 
$\widetilde{\rm  G}_{\rm N}(t=0)$ cannot be directly constrained from the running of $g$, which scales with $k$ as $(k/M_s)^{26}$ as  follows from 
the non-perturbative renormalisation flow of $g^2(t)$.  We can therefore assume it also finite. 

Thus, from \eqref{gGrun}, we arrive that at the non-trivial infrared fixed point of the RG flow (see \eqref{fphermit2}), in which $\lim_{t \to -\infty} g(t)=g^\star < \infty$, finite, we obtain for the 
Newton's gravitational parameter in the IR region of this flow:
\begin{align}\label{irGrun}
\lim_{t \to -\infty}\widetilde{\rm G}_{\rm N} (t ) = \lim_{t \to -\infty} \frac{g^2(t=0)}{g^{\star \,2}}\, \widetilde{\rm G}_{\rm N} (t=0) \, e^{2t} =  0\,, \nonumber
\end{align} 
indicating that the gravitational constant at the singularity goes to zero.  The finite value $\widetilde{\rm G}_{\rm N} (t=0)$ is identified with the Newton's constant in the UV regime (short distances). 
In \eqref{sea3} the axions couple  to both gravitons and gauge fields;  the presence of gravitons is expected to modify the axion self energy, and through this their nonperturbative running. 
However, the topological (metric independent) nature of the Pontryagin density term of \eqref{sea3},
$b\, F\, \widetilde F$, is expected not to alter the singularity structure of the axion coupling, thus leading to the aforementioned properties of the running gravitational coupling, in particular its vanishing in the infrared.

Transitioning to the non-Hermitian CS theory, is achieved by replacing $g$ with $ig$, which, in the context of our string inspired theory \eqref{eq:AxEl}, \eqref{axionstringcoupl}, is equivalent to
the replacement 
of the gravitational constant in the effective action \eqref{sea3} by a purely imaginary one,
\begin{equation}\label{kappacomplex}
\kappa \,\, \rightarrow \,\, i \,\tilde \kappa, \quad \tilde \kappa \in \mathbb R\,.
\end{equation}

Under this very restrictive complexification  of the gravitational coupling, the effective action becomes a \cPT-symmetric nonHermitian 
action,\footnote{Our approach does not consider issues related to the conjecture that there are no  global symmetries in quantum gravity~\cite{Harlow:2022ich}.} describing repulsive (anti)gravity dynamics:\footnote{As already mentioned, the presence of imaginary CS terms implies the loss of periodicity of the effective axion potential, in contrast to the Hermitian case ({\it cf. }\eqref{indexx}, \eqref{axion}). In the nonHermitian case, one may have though dynamical generation of the axion mass, as in the scenarios exploited in \cite{R3.13,Alexandre:2020tba,R3.14,R3a,R3c,Mavromatos:2023bdx}. We hope to come back to such scenarios in a future publication.}
\begin{align}\label{anti}
  S^{\rm eff \, anti}_{\rm B}&=\; \int d^{4}x\sqrt{-g}\Big[ \dfrac{1}{2\widetilde \kappa^{2}}\, R + \frac{1}{2}\, \partial_\mu b \, \partial^\mu b \nonumber \\
& -i\,\sqrt{\frac{2}{3}} \, \frac{\alpha^\prime}{96\, \widetilde \kappa} \, b(x) \, \Big(R_{\mu\nu\rho\sigma}\, \widetilde R^{\mu\nu\rho\sigma} - \mathbf F_{\mu\nu}\, \widetilde{\mathbf F}^{\mu\nu}\Big) + \dots \Big],
\end{align}
In the non-Hermitian theory, the running of the gravitational coupling with $t$ also follows from \eqref{axionstringcoupl}, and we have the following relation
connecting the values of the couplings at the trivial IR fixed point with their values in the UV one of the pertinent RG flow:
\begin{align}\label{trivialIR}
\lim_{t \to -\infty}g^2(t ) \, {\widetilde{\rm G}}_{\rm N} (t \to -\infty)  
= \lim_{t \to -\infty} \frac{g^2(t=0)}{g^{\star \,2}}\, \widetilde{\rm G}_{\rm N} (t=0) \, e^{2t} =0\,.
\end{align}
But at the trivial IR fixed point $\lim_{t \to -\infty}g^2(t) \to 0$ (see fig.~\ref{fig:betagplot}), while at the non-trivial UV fixed point $g^2(t=0) = g^{\star\, 2} < \infty$, a fixed finite value. With $\widetilde{\rm G}_{\rm N} (t=0)$ finite, as discussed above, this leaves $ \widetilde{\rm G}_{\rm N} (t \to -\infty) $ also undetermined, and we assign it a finite value, which is that of a Newton's gravitational constant at large scales.

This scenario may lead to a 
reinterpretation of the dark-energy sector of the universe observed at large distances. The issue is to observe the possibility that {\it unbroken} $\cPT$ symmetry is a {\it low energy} symmetry of the real world. In this sense repulsive gravity would correspond to a low-energy (large distance) phase of the universe, which would be described effectively by the non-Hermitian \cPT-symmetric gravitational effective action \eqref{anti}
of an underlying string theory.
In summary, upon assuming the phase transition at $k_{\rm sing}$ to a phase whose dynamics is described by the non-Hermitian theory for $k < k_{\rm sing}$, one can connect smoothly  a zero gravitational constant ${\rm G}(k_{\rm sing})=0$ to a 
{\it negative}, non-zero and \emph{finite} Netwon's constant ${\rm G}(m_{\rm IR}) < 0$. 
This behaviour that characterises our string-inspired Chern-Simons gravity model is different from the standard RG analysis in asymptotic safety model ({\it cf.} Eqs.~\eqref{rgG}, \eqref{IRgr}). 
In our analysis, loop corrections in the gravity sector have not been considered, and so we can only provide arguments on this behavior for gravity in this CS model, but not a proof.

The complete answer on the running of Newton's gravitational parameter in our \cPT-symmetric-nonHermitian-gravity approach can only be determined when the full quantum dynamics of the gravity sector in the model is taken into account.  
Nonetheless, we make the important remark that the topological nature of the gauge Pontryagin density (but not of the Hirzebruch term in \eqref{sea3}, whose variation with respect to the metric field yields the Cotton tensor~\cite{jackiw,Alexander:2009tp}) is encouraging for the maintenance of the crucial singularity structure of the axion coupling in such cases.
Further analysis in the gravitational sector is necessary to fully understand this behaviour, but this is beyond the scope of the current article.

\section{Conclusions and Outlook \label{sec:concl}}

This article presents a novel perspective on repulsive gravity at large cosmological distances, potentially explaining the observed accelerated expansion of the universe. 
It contrasts the conventional interpretation of this accelerated expansion in terms of a positive cosmological constant or dark energy,  with the proposal of a \cPT-symmetric phase of gravity.
The gravitational system, which leads to  this conjecture, is a string-inspired Chern-Simons gravity, involving the coupling of axions to anomalous Chern-Simons gauge and gravitational terms. Thus the conjecture may be specific to such string-inspired systems.

Our conjecture is based on having a  running Newton's coupling (i.e. the inverse of the square of the reduced Planck mass), which in the ultraviolet regime exhibits an asymptotic safety behavior, \eq{rgG},  with the running coupling tending to a nontrivial UV fixed-point value as the scale $k^2/m^2_{\rm IR}\to\infty$. In the IR, however, the running leads to the beta function of the Hermitian system  developing a singularity at a running RG scale $k_{\rm sing} \sim 1.01 \, m_{\rm IR} \to 0$. There is a relation  \eqref{axionstringcoupl} between the axion couplings in the string effective action \eqref{sea3} and the coupling in field theory for  massless  axion electrodynamics, which is the limit of the string-effective gravitational theory for flat-spacetime and Abelian-gauge-group. We refer to this connection as an embedding. The IR singularity corresponds to a phase transition in axion electrodynamics.
The Hermitian version of the embedded model exhibits an RG flow ({\it cf.} fig.~\ref{fig:betagplot})
characterised by a singularity of the beta function \eqref{gvalaxqed} in the IR region of the running RG scale \eqref{ksing}. We propose that such a region corresponds to a phase transition of the system, such that the theory that describes the singularity, as well as the region $k < k_{\rm sing}$, is the non-Hermitian model obtained by $g \to i\tilde g$, $\tilde g \in \mathbb R$. This theory exhibits a smooth behavior in the RG running from $k_{\rm sing}$ of the Hermitian model to the trivial IR fixed point at $\tilde g=0$ ({\it cf.} fig.~\ref{fig:Beta fn}).
Bypassing the singular behavior is a highly model-dependent issue. In most of the approaches so far, the attractive nature of gravity is maintained when the IR behavior of gravity is regulated \eqref{IRgr}. In this work we have given a prescription for avoiding the IR singular behavior of the Newton's coupling, as the running RG scale $k^2/m^2_{\rm IR} \to 0^+$, within the context of the string-inspired gravity model with axions. The presence of the anomalous topological couplings are crucial for inducing the repulsive $\cPT$ phase.
In our case, the embedded nonHermitian axion electrodynamics theory leads in the IR region of the RG running,  to a repulsive phase of gravity upon using \eqref{kappacomplex}, in which the axion coupling $g$ is related to the gravitational scales of string theory \eqref{axionstringcoupl}. This gives a prescription for  $\cPT$-symmetric version of repulsive gravity at large (cosmological) scales. 
For a more complete  justification of the conjecture of the repulsive phase of gravity, we need a proper (non-perturbative) quantisation of the effective gravitational theory obtained from strings. This is very challenging. 

Finally, we stress that ordinary quantized Einstein gravity does not appear to exhibit repulsive behaviour \cite{Reuter_Saueressig_2019}. \color{black} It is only after the addition of a (dominant) positive cosmological constant (de-Sitter type) term, that the effective force appears repulsive.\footnote{\color{black} Although, even this interpretation has been questioned in \cite{bernabeu}, as mentioned previously in section \ref{sec:asymptsafety}.\color{black}}
\color{black}
We only see this effect in anomalous string-inspired gravity with gravitational axions \eq{sea3} (\cite{str1, str2}). Such features do characterize strings, but one can {\it also} postulate local gravity theories that share them, most notably modified gravity theories with torsion~\cite{Cartan:1938ph,hehl,shapiro,Hammond:2002rm,PhysRevD.104.084067,iorio}, known to be associated with propagating axionic degrees of freedom. Apart from the possible connection with repulsive gravity, our scenario of the emergence of \cPT ~symmetry  is also connected to a change in the generalised symmetries \cite{Bhardwaj:2023kri} of  axionic electrodynamics. This deserves closer attention {\it per se} which we hope to consider in a future publication.

\color{black} As mentioned in the introduction (subsection \ref{sec:Theme}), we contrast our approach with the study  of quantum field theories with couplings for repulsive and attractive interactions, and the  
 renormalization group flow of the  couplings. The latter flow  is not in general associated with the existence of a singular point of the renormalization-group flow, unlike our case. 
In our case the change of sign of the gravitational interaction is \emph{induced} by the existence of singular points of (the nonperturbative) renormalization-group flows of the axion-electrodynamics part of the curved-space string-inspired CS effective action in \eqref{sea3}; this flow necessitates a non-trivial jump to the \cPT ~symmetric framework, in order to bypass such a singularity. This has been interpreted in our work as indicating a phase transition of the system to a repulsive gravity phase.

From a phenomenological view point, our gravitational theory \eqref{sea3}, being a CS gravity~\cite{jackiw,Alexander:2009tp}, has features distinguishing it from ordinary gravity.
One of them is the \emph{birefringent} behaviour of gravitational waves (GW). Indeed, when GW are produced through non-spherically symmetric coalescence of, say, primordial black
holes or collisons of domain walls~\cite{Mavromatos:2020kzj}, the gravitational CS term becomes non-trivial because different helicities (left, right) of the tensor (metric) perturbations propagate in a different way. This difference of the wave equations for the left and right -
handed polarizations arises from  the (non-trivial)  metric variation of the gravitational CS term~\cite{jackiw}, which, in turn, modifies the gravitational
equations leading to gravitational wave \emph{birefringence} of cosmological origin~\cite{Alexander:2004us,Lyth:2005jf,Alexander:2004wk}.\color{black}\footnote{In our model \eqref{sea3}, the (Parity-violating) gravitational CS term couples to pseudoscalar (axion-like) fields $b(t, \vec x) = - b(t, -\vec x)$, hence the relevant interaction respects Parity. In cases in which there are scalar fields (e.g. dilatons) that couple to the gravitational CS term,  there is in addition cosmic Parity violation, which can leave its imprints on cosmic microwave background radiation~\cite{Lue:1998mq} in the early universe.}
\color{black}

 \color{black} However, for our purposes the most important  phenomenological consequence of our approach is the effect on the associated phase transition to a $\cPT$-symmetric phase of gravity at late epochs in the evolution of the universe. The latter could leave important and novel imprints on the associated GW, in analogy, e.g., with the effects of phase transitions in the early universe (e.g. QCD, first-order electroweak, topological defects {\it etc.}) on the resultant 
 GW profiles~\cite{Caprini:2015zlo,Caprini:2019egz,Kosowsky:1992rz,Grojean:2006bp,Cutting:2020nla}. 

Our conjectural model of cosmic acceleration is too early  in its development  to predict exactly the pertinent features in the profiles of the GW induced by the $\cPT$ phase transition, given that its nature is still under investigation. Nonetheless, we believe that this is a research avenue worthy of further study, that could elucidate potential phenomenological/observational tests of this model for the cosmic acceleration. We hope to be able to report on such issues in future works.

\color{black}

\vskip 2cm

\section*{Acknowledgements} 
We thank Carl M. Bender for discussions on the $\cPT$-symmetric quantum mechanical anharmonic oscillator. The work of NEM and SS is supported in part by the UK Science and Technology Facilities research Council (STFC) and UK Engineering and Physical Sciences Research Council (EPSRC) under the research grants  ST/X000753/1 and  EP/V002821/1, respectively. NEM also acknowledges participation in the COST Association Action CA21136 “Addressing observational tensions in cosmology with systematics and fundamental physics (CosmoVerse)”.

\section*{References}
\medskip

\bibliographystyle{apsrev4-1}

\bibliography{CSreview_June.bib}

\end{document}